\documentclass[12pt,oneside,reqno]{amsart}
\usepackage{amstext}
\usepackage{amssymb}
\usepackage[letterpaper]{geometry}
\usepackage{color}
\usepackage{float}
\usepackage{units}
\usepackage{graphicx}
\usepackage{wasysym}
\usepackage{natbib}
\usepackage[colorinlistoftodos]{todonotes}

\makeatletter

\providecommand\textquotedblplain{%
  \bgroup\addfontfeatures{Mapping=}\char34\egroup}
\providecommand{\tabularnewline}{\\}

\date{}

\usepackage{caption}
\usepackage{wasysym}
\usepackage{accents}

\makeatother

\begin{document}
\title{On the Problem of Modeling the Boat Wake Climate; the Florida Intracoastal
Waterway}

\author{Carola Forlini\textsuperscript{1}, Rizwan Qayyum\textsuperscript{1},
Matt Malej\textsuperscript{2}, Michael-Angelo Y.-H. Lam\textsuperscript{2},
Fengyan Shi\textsuperscript{3}, Christine Angelini\textsuperscript{1},
Alex Sheremet\textsuperscript{1}}

\curraddr{1. Engineering School of Sustainable Infrastructure and Environment,
University of Florida\\
2. U. S. Army Engineer Research and Development Center, Coastal and
Hydraulics Laboratory\\
3. Center for Applied Coastal Research, Department of Civil and Environmental
Engineering, University of Delaware}

\maketitle
\begin{abstract} 
The impact of boat traffic on the health of coastal ecosystems is a multi-scale process: from minutes (individual wakes), to days (tidal modulation of sediment transport), to seasons and years (traffic is seasonal). A considerable numerical effort, notwithstanding the value of a boat-by-boat numerical modeling approach, is questionable, because of the practical impossibility of specifying the exact type and navigation characteristics for every boat comprising the traffic at any given time. Here, we propose a statistical-mechanics description of the traffic using a joint probability density of the wake population in some characteristic parameter space.  We attempt to answer two basic questions: (1) what is the relevant parameter space and (2) how should a numerical model be tested for a wake population? We describe the linear and nonlinear characteristics of wakes observed in the Florida Intracoastal Waters. Adopting provisionally a two-dimensional parameter space (depth- and length-based Froude numbers) we conduct  numerical simulations using the open-source FUNWAVE-TVD Boussinesq model. The model performance is excellent for weakly-dispersive, completely specified wakes (e.g., the analytical linear wakes), and also for the range of Froude numbers observed in the field, or for large container ships generating relatively long waves. The model is challenged by the short waves generated by small, slow boats. However, simulations suggest that the problem is confined to the deeper water domain and linear evolution. Nonlinear wake shoaling, essential for modeling wake-induced sediment transport and wake impact on the environment, is described well.
\end{abstract}

\section{Introduction}

\subsection{Motivation}

The boat traffic associated with the accelerating urbanization of coastal areas has a significant impact on the health of  coastal ecosystems. In otherwise typically quiescent intracoastal and estuarine waterways, the transient wave activity created by boat traffic generates intermittent bursts of higher flow velocities, accelerations, and stresses on the bed, vegetation, and both bivalve and coral reefs. Although boat-wakes may have relatively small heights in deep water, they grow, steepen, and break as they shoal on sloping banks/beaches much like ocean waves, with similar sediment-transport effects. Fine grained sediment is mobilized and transported, increasing water turbidity as well as exposing vegetation roots, eroding reefs, and destabilizing the shoreline. Despite growing awareness and concern \citep[e.g.,][]{Sidman2005,Scarpa2019}, the effect of the boat-wake climate on coastal ecosystems is not well understood, leaving management agencies with few tools for either evaluating their degradation or for estimating the effectiveness of possible mitigation measures. 

Quantifying the long term impact of boat wakes\footnote{In general, the term "wake" is used to denote the trace left on the water by the moving perturbation (ship/boat), i.e., it refers to both turbulent and wave components of the trace. Because here we ignore the turbulent component, with a slight abuse of terminology, we use the terms "ship waves"  and "ship wakes" to denote only the wave component.} is a complex problem that requires an effective, multiple-scale characterization of boat traffic, local hydrodynamics, as well as the local ecosystem. Depending on the site, ship traffic may comprise a complex spectrum of vessel types, from commercial freight, fishing and passenger ships, to recreational boats of various sizes \citep[e.g.,][]{Herbert2018}.  
Because the generated waves reflect the diversity of shapes and navigation characteristics of the generating vessels, an accurate quantitative description of all of the vessels that constitute the traffic, including such parameters as vessel type, shape, speed, pressure footprint, direction of navigation, two-dimensional path curvature, and perhaps others, is essential for a representative description of the wake field at any given time. While such a characterization of the traffic is conceivable, in practice it is almost impossible to obtain. 

The most challenging physical aspect of the wake-climate impact on the coastal and nearshore ecosystems is the multi-scale process of accumulation of small effects over long periods of time. Individual wakes cause a short time-scale perturbation on the order of minutes. If one accepts the conceptual ``stir and transport'' model, with the boat wakes intermittently stirring the bed sediment and currents (tidal, or of other origin) transporting it away, then the relation between peak traffic (and its characteristics) and current patterns is paramount – acting at a time scale on the order of hours to days. Longer time-scale modulations include atmospheric perturbations and storms that may affect both hydrodynamic circulation and traffic patterns at time scales on the order of weeks, as well as seasonal changes in the hydrodynamics and ship traffic composition and intensity. Adding an entirely new dimension of complexity, the multiple-scale response of the living ecosystems that interact with this wake climate is perhaps the least understood. 

\subsection{The wake climate problem\label{subsec: climate}}

Because of its multi-scale nature, attempting to model the long-term ecological impacts of boat wakes by modeling each individual wake seems impractical. Since wakes are nonstationary, anisotropic, and inhomogeneous wave fields, they require phase-resolving, time-domain models, that incorporate natively processes such as nonlinear wave transformation and breaking, wave-current interaction, swash zone dynamics, and wave-sediment interaction (powerful statistical tools developed for wind-wave processes, e.g., \citealp{Cavaleri2007}, are not applicable). Phase-resolving wave modeling on large computational domains can be time consuming. Moreover, the numerical effort of modeling every wake that possibly impacts the environment over weeks and months would be operationally difficult to replicate with a sufficient level of accuracy to be meaningful for decision making or advancing understanding of ship impacts on coastal ecosystems. An additional consideration is the accumulation of errors in the long-term description of a wave climate at hand.  On a more fundamental level, this approach  may not be operationally feasible, as the entire vessel-passage climate with reasonable accuracy and navigation parameters is needed for modeling each individual wake. 

However, these types of difficulties are in fact common to physical systems comprising a large number of components (e.g., a gas), and follow a similar blueprint: the equations governing the dynamics of each component are available, but cannot be integrated directly to obtain the evolution of the system as a whole, because 1) there are just too many components (equations), and 2) the initial conditions are not known (e.g., the position and speed of each gas molecule). The case of the wake climate, while numerical simulations have the ability to handle multiple boat wakes simultaneously, the numerical effort remains significant because of the long characteristic  time scale of the wake climate impacts. Moreover, in practical applications the the exact initial state of each boat is difficult to specify in practice (see also section \ref{sec: obs}). An alternative approach is suggested by statistical mechanics. The key step forward is to recognize that the goal is not to provide an accurate description of the dynamic of each component, but to model the macroscopic (long time-scales) dynamics of the system. While the microscopic (boat) scale dynamics have a very large number of degrees of freedom, macroscopic dynamics are low-dimensional. For example, an ideal gas composed of identical molecules has $6N$ microscopic degrees of freedom, where $N$ is the number of molecules, but only 2 macroscopic degrees of freedom -- volume and entropy. This implies that a macroscopic state corresponds to a very large number of microscopic configurations: the exact state of any given gas molecule is not relevant; what is relevant is how individual molecules distribute, say, by their kinetic energy. The problem of solving the $6N$ microscopic dynamical equations is then replaced by the problem of describing the evolution of probability density functions. 

Here, we propose that a statistical-mechanics approach is possible for describing the impact of the boat traffic on the environment. The macroscopic system is the \emph{wake climate}, defined as a large population of wakes impacting a given area on a long time-scale. The system components are individual wakes. In switching scales from microscopic to macroscopic, we make the fundamental assumption that the exact details of the individual wakes  do not matter at  macroscopic scale, largely erased by some averaging process. What matters is the distribution of wakes,  described by joint probability densities in wake-parameter space. Wake characteristic  parameters might include Froude numbers, boat form factors,  draft, temporal and spatial distributions and others. Analytical averaging techniques would probably be unfeasible, but instead of modeling individual wakes one could perform a Monte-Carlo simulation based on the identified \emph{classes of wakes},  weighted by their probability density. 

Such an approach poses at this time two basic challenges. First, we do not know what is the optimal parameter set (space) to characterize wakes for this purpose. Despite recent efforts to understand the structure of both individual wakes \citep[e.g.,][]{Sheremet2012,Didenkulova2013,Torsvik2015,Pethiyagoda2017,Herbert2018} and the wake population, it is not yet clear what parameters are relevant for quantifying the wake climate, and thus how to describe wakes as a population. Obvious candidates are Froude numbers defined by a linear formulation of the boat-wake problem (see below). However,  because the ultimate problem is not wake evolution but its average impact on the environment (i.e. channel bed, subtidal and intertidal ecosystems), one has to assume that other parameters are also important, such as those related to  different  wake components (divergent and transversal waves), the characteristic wave length, energy and energy flux (or related parameters maximum height), as well as some characterization of the nonlinear shoaling evolution of the wake, such as breaking depth. The frequency of occurrence of wakes,  their spatial distribution, and their relation with other processes such as tides and local currents should  also be important.   One should also note that, obviously, not all wake climates are the same. For example, the wake climate near a busy commercial-vessel port such as Galveston/Houston, Texas is significantly different from that of intracoastal waters in Florida, where traffic is composed mainly of recreational boats and watercraft. From the numerical modeling perspective, local conditions may play an essential role in choosing the type of model to be used. For example, Boussinesq models are adequate if the wake-parameter space is essentially in the weak dispersion range, but might not be useful otherwise.  This challenge is the focus of this study.

The second challenge is modeling the interaction of wake classes with local circulation, and their impact on the ecosystem. Once a numerical model is identified, a possible approach to simulating the wake climate might involve the following steps: 1) estimate the statistics of the time evolution of the boat traffic at a given location; 2) use these data to estimate time and spatial evolution of the wake distribution in a defined parameter space; 3) model the interaction and impact of each wake class with the local circulation conditions using accurate models of wake dynamics; 4) estimate the time evolution of the wake impact on the ecosystem based on the characteristic impact of each class and its weight in the population of wakes (assuming, say, that wake realizations are independent). We will not discuss further this aspect in this study. 

\subsection{Numerical modeling}

As noted above, the choice of numerical model is essential for the statistical approach, because the model should be capable to simulate well all the wake classes observed at a given location. On the one hand, this choice is complicated by field observations show widely-varying wake structures, with wake elements having different scales and possibly different sediment transport impacts \citep{Sheremet2012,Didenkulova2013,Torsvik2015}. For example, the drawdown wake of a slow-moving, slender vessel with high draft/channel-depth ratio may produce more damage than a faster moving cruiser wake, which has a quite different structure \citep[e.g.,][]{Scarpa2019}. On the other hand, the family of numerical wave models that can be applied to vessel-generated waves is not large. Finite-element Computational Fluid Dynamics (CFD) models/packages (e.g., OpenFOAM) allow for detailed modeling of vessel hull type and the resulting wakes, but the computationally intensive numerical effort is prohibitive for operational use.  Because the sites impacted by wakes are typically in shallow water, Boussinesq-type models are obvious candidates \citep[e.g.,][]{Nwogu2001,Madsen2003,Shi2012,Malej2015,Shi2016,David2017,Malej2019}; see also reviews by \citet{Kirby2016,Brocchini2013}.  They  have become widespread with the advent of high-performance parallel computing systems, and can describe accurately wave propagation, transformation and other nearshore processes over many physical length and temporal scales in an operational setting \citep{Malej2015}. 

Here, we investigate the applicability of FUNWAVE-TVD \citep{Shi2012} to modeling boat wakes in the Intracoastal Waterway in northeast Florida. The model has a long history of applications and testing (wave shoaling, refraction and diffraction, breaking, wave-induced nearshore circulation, tsunami waves and coastal inundation, as well as laboratory ship-wake generation and propagation \citealt{Shi2012b,kirby2013,Tehranirad2011,Lynett2017,Shi2018}). Moreover, modules have been developed recently for vessel-generated waves by different pressure distributions as well as sediment transport effects, including morphological changes \citep{Malej2019}. While other formulations, such as the Boussinesq Ocean and Surf Zone (BOSZ) model \citep{Roeber2012} have included and tested recently ship-wave modules with reasonable skill \citep[e.g.,][]{David2017}, the FUNWAVE-TVD model is given preference here because of its scalable powerful MPI implementation and its flexible choice of ship-hull representation.

In this study, we analyze field observations with the goal of identifying and classifying observed wakes, and examine their statistical distribution. Based on a rather simple linear wake theory, we formulate the characteristic wake-parameter space in terms of Froude numbers, and use it to investigate the ability of the FUNWAVE-TVD  to describe different types of  wakes.  The methods are described briefly in section \ref{sec: methods};  we discuss wake classes observed in the Florida Intracoastal Waterway in section \ref{sec: obs}; the numerical tests are presented in section \ref{sec: num}; some conclusions and implications of this study are discussed in section \ref{sec: disc}.

\section{Methods\label{sec: methods}}

\subsection{Analysis of field data} 

Pressure measurements were converted to free surface elevation by transforming the time series to the Fourier domain, multiplying the complex amplitudes by the factor $\beta\frac{\cosh kd}{\cosh kh}$ and transforming back to the time domain (e.g., \citealp{Sheremet2012,Torsvik2015}). Here, the wave number $k$ is related to the radian frequency $\omega=2\pi f$ through the dispersion relationship $\omega^{2}=gk\tanh kd$, where $f$ is the frequency, $d$ is the local depth, and $h$ is the height of the instrument above the bed. To avoid amplifying noise in the high-frequency tail of the Fourier spectrum, a smooth cutoff in the neighborhood of 1 Hz was introduced in the amplification factor using the sigmoid function $\beta(f)=0.5\left(1-\tanh\frac{f-f_{0}}{w}\right)$, where $f_{0}$ is the inflection point of the sigmoid and $w$ the width of the transition. The values used here, $w=5$ Hz$^{-1}$ and $f_{0}=0.5$ Hz, were chosen to minimize amplification artifacts.

Because boat wakes are highly non-stationary (the characteristic times of amplitude and phase modulation, both linear and nonlinear,  are similar to the characteristic period;  for example,  boat wakes shown below have a duration in the order of 0.5-1 min, with a typical period is on the order of 2-4 s.), analysis methods that assume stationarity and average over time do not work well on wakes. As a result, standard spectral and bispectral  have to be replaced by non-averaged estimators that have a more limited scope. The basic tool is the short-time, or Windowed Fourier Transform (WFT) and its inverse, which may be introduced formally as decomposition/reconstruction of a real function $g(t)$, $t\in\mathbb{R}$ using a set of elementary functions
\begin{equation}
\psi_{f,\tau}(t)=w_{\tau}(t)e^{2\pi ift},\;\int_{-\infty}^{\infty}w_{\tau}(t)dt=1\label{eq: basis}
\end{equation}
where $w_{\tau}(t)$ is a window centered at $\tau$, a real function of finite support or decaying fast enough away from its center, and $f,\tau\in\mathbb{R}$, i.e., 
\begin{equation}
G_{\tau}(f)=\int g(t)\psi_{f,\tau}^{*}(t)dt,\;g(t)=\iint_{-\infty}^{\infty}G_{\tau}(f)\psi_{f,\tau}(t)dfd\tau.\label{eq: I/WFT}
\end{equation}
The function $G$ is usually called ``periodogram''. In a way similar to the cross-spectrum, if $g(t)$ and $h(t)$ are two time series with periodograms $G_{\tau}(f)$ and $H_{\tau}(f)$, we define the cross-spectrogram $S^{gh}$ of $g$ and $h$ as
\begin{equation}
S_{\tau}^{gh}(f)=G_{\tau}(f)H_{\tau}^{*}(f),\label{eq: xsptg}
\end{equation}
where the asterisk denotes complex conjugation. For a single time series $g$, $S_{\tau}^{gg}$ is called the spectrogram. The mean frequency associated with a defined wake component is calculated as
\begin{equation}
f_{mean}=\frac{\iint_{\mathcal{D}}fS_{\tau}^{gg}(f)d\tau df}{\iint_{\mathcal{D}}S_{\tau}^{gg}(f)d\tau df},\label{eq: tf mean}
\end{equation}
where $\mathcal{D}$ is the domain the time-frequency space occupied by the wake component. 

The nonstationarity of the wakes affects also higher-order spectral estimators, commonly used for evaluating the nonlinear characteristics of a wave. To examine the phase coupling between different components of the wake, we generalize the spectrogram idea to bispectra by defining the biperiodogram  of time series $g$ as
\begin{align}
B_{\tau}^{g}\left(f_{1},f_{2}\right) & =G_{\tau}\left(f_{1}\right)G_{\tau}\left(f_{2}\right)G_{\tau}^{*}\left(f_{1}+f_{2}\right)\label{eq: bpdg}
\end{align}
The bispectrum estimate is essentially a normalized  biperiodogram averaged over realizations (in practice, over all windows $w_{\tau}$), for example,
\begin{align}
\mathcal{B}^{g}\left(f_{1},f_{2}\right) & =\frac{\left\langle B_{\tau}^{g}\left(f_{1},f_{2}\right)\right\rangle }{\left(\left\langle S_{\tau}^{gg}(f_{1})\right\rangle \left\langle S_{\tau}^{gg}(f_{2})\right\rangle \left\langle S_{\tau}^{gg}(f_{1}+f_{2})\right\rangle \right)^{\nicefrac{1}{2}}}.\label{eq: bspt-1}
\end{align}
(other normalizations have been proposed, e.g., \citealp{Haubrich1965,Elgar1985}). The bispectrum is statistically zero if the Fourier coefficients are statistically independent, therefore the bispectrum is a measure of the coupling between different frequency components of $g$. The bispectrum is related to third-order ``global'' measures of departure from Gaussianity,
\begin{equation}
\left(\sigma^{g}\right)^{-3}\iint\mathcal{B}^{g}\left(f_{1},f_{2}\right)df_{1}df_{2}=\gamma^{g}+i\alpha^{g},\label{eq: B=00003DS+iA}
\end{equation}
where $\gamma^{g}$ is the skewness, and $\alpha^{g}$ is the asymmetry of time series $g$ \citep{MasudaKuo81,Elgar1987}.  Although the definition of the biperiodogram parallels that of the periodogram, its interpretation is less obvious, because averaging and normalization operations applied to the bispectrum to bring out time-consistent cross-frequency phase relations, are not possible in case of a wake.  By itself, the periodogram \ref{eq: bpdg} only shows local phase relations, and without a normalization, it is also dominated by the frequency components that have the largest amplitude. However, because the wake has typically a narrow instantaneous power distribution, the position of biperodogram peaks may still be interpreted as an indicator of phase relations.

 Below, we use the real and the imaginary part of the biperiodogram to examine the time evolution of cross-component phase relations. Based on equation \ref{eq: B=00003DS+iA}, following \citet{MasudaKuo81,Elgar1987} we interpret the real part of the biperiodogram $\Re\left\{ B_{\tau}^{g}\left(f_{1},f_{2}\right)\right\} $ as a measure of the {\em local} skewness, and the imaginary part $\Im\left\{ B_{\tau}^{g}\left(f_{1},f_{2}\right)\right\} $ as a measure of the  {\em local} asymmetry of the signal (note that standard skewness and asymmetry definitions are global statistical parameters, and implicitly assume stationarity). This interpretation is supported by the local aspect of the time series. 

In all the applications shown, numerical estimates of these quantities are obtained using the discrete Fourier Transform. (details of numerical implementation and the use discrete Fourier Transform can be found in, e.g., \citealp{Briggs1995}). The data analysis procedures were implemented in the Matlab® environment using available Matlab® toolboxes. For the direct and inverse WFT we used the Matlab® codes for Short-Time Fourier Transform (STFT) and its inverse (ISFTF) distributed by \citet{Zhivomirov2019}. The typical window length used for the WFT was $2^{8}$ points, equivalent to a 32-s time window sampled at 8 Hz. 

\subsection{ Basic parameter space: an analytical linear model }

\begin{figure}
\begin{centering}
\includegraphics[width=0.9\textwidth]{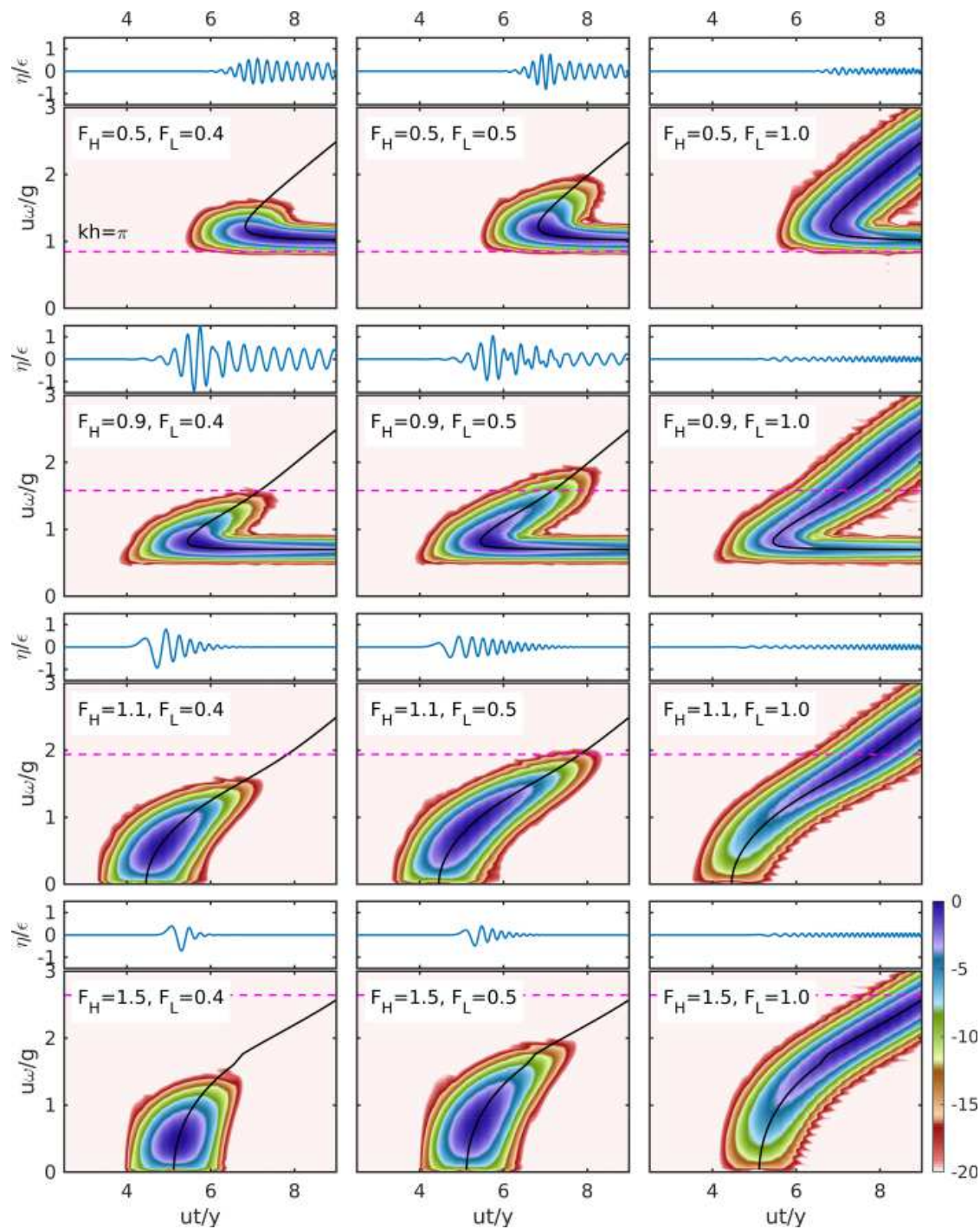}
\par\end{centering}
\caption{Examples of wakes generated by a Gaussian pressure distribution moving with constant speed over finite-depth water, classified by depth- and length-based Froude numbers $F_{H}$ and $F_{L}$. The free surface elevation and corresponding spectrogram are shown for each pair $(F_{H},F_{L})$ of characteristic Froude numbers. Spectrograms are normalized to unit global maximum, and represented $\log_{10}$ scale. The continuous black line in the spectrogram panels represents \citet{Pethiyagoda2017} ``dispersion'' curve. The chirp branch (frequency increasing in time) corresponds to the divergent wake, and is generated in both sub- and super-critical wakes. The monochromatic (frequency constant in time) branch corresponds to the transversal wake and is generated only in subcritical wakes.  The dashed line corresponds to $kh=\pi$. The analytical model used to generate this figure is based on \citet{Pethiyagoda2018} using Matlab®. \label{fig: Gaussian wakes}}
\end{figure}
The fundamental Kelvin wake solution \citep{Kelvin1887} of the linear wave equations is well understood \citep{Havelock1909,Havelock1918,Lamb1997,Wehausen1960,Olver2010}. The scaled equations for the irrotational, incompressible flow generated by a pressure distribution moving with speed $U$ in the $x_{1}$ are \citep{Pethiyagoda2017}
\begin{gather}
\phi_{xx}+\phi_{zz}=0\;\text{in}\;\ensuremath{z<\eta,}\label{eq: Laplace}
\end{gather}
with boundary conditions
\begin{gather}
\left[\frac{1}{2}\left(\phi_{x}\right)^{2}+\eta+\epsilon p\right]_{z=\eta}=\frac{1}{2},\;\left[\phi_{x}\cdot\eta_{x}-\phi_{z}\right]_{z=\eta}=0,\;\left(\phi_{z}\right)_{z=F_{H}^{-2}},\;\begin{aligned}\phi & \rightarrow0\end{aligned}
\text{as \ensuremath{x\rightarrow-\infty}},\label{eq: BC}
\end{gather}
where $\phi(x,z)$ is the velocity potential, with $x=(x_{1},x_{2})$ the horizontal, two-dimensional coordinate; $\eta(x,y)$ is the free-surface elevation; subscripts $x$ and $z$ denote differentiation; $\epsilon p\left(x,y\right)$ is the pressure distribution, with $\epsilon=D/\rho U^{2}$, with the ``draft'' $D$ denoting the maximum pressure value. The parameter $F_{H}=U/\sqrt{gh}$ is the depth-based Froude number, $\rho$ is the water density, and $g$ is the gravitational acceleration. In equations \ref{eq: Laplace}-\ref{eq: BC}, the scaled speed of the ship is 1 and the unit of length is $\frac{U^{2}}{g}$. 

In the limit $\epsilon\to0$ equations \ref{eq: Laplace}-\ref{eq: BC} are linear, and for finite-depth have the solution  \citep[e.g.,][]{Wehausen1960} 
\begin{align}
\eta(x)= & -\epsilon p\left(x\right)+\frac{\epsilon}{2\pi^{2}}\int_{-\pi/2}^{\pi/2}d\theta\int_{0}^{\infty}dk\frac{k^{2}p\left(k,\theta\right)\cos\left[k\left(|x|\cos\theta+\sin\theta\right)\right]}{k-\sec^{2}\theta\tanh\left(k/F_{H}^{2}\right)}\nonumber \\
 & -\frac{2\epsilon F_{H}^{2}H(x)}{\pi}\int_{\theta_{0}}^{\pi/2}d\theta\frac{k_{0}^{2}p\left(k_{0},\theta\right)\sin\left(k_{0}x\cos\theta\right)\cos\left(k_{0}y\sin\theta\right)}{F^2_{H}-\sec^{2}\theta\text{sech\ensuremath{^{2}\left(k_{0}\right)}}}
 \label{eq: linear solution}
\end{align}

where the integration over $k$ is taken below the pole $k=k_{0}$,
with $k_{0}$ the real positive root of 
\[
k-\sec^{2}\theta\tanh\left(\frac{k}{F^2_H}\right)=0,\;\theta_{0}<\theta<\frac{\pi}{2},
\]
$p(k,\theta)$ is the Fourier transform of the pressure distribution
$p(x,y)$ representing the footprint of the moving ship, $H(x)$ is
the Heaviside function, and $\theta_{0}=0$ if $F_{H}<1$ and $\theta_{0}=\arccos F_{H}^{-1}$
if $F_{H}>1$. The far-field wake is dominated by the second integral, because (by the usual stationary-phase argument) the double integral in equation \ref{eq: linear solution} rapidly tends to zero far away from the pressure distribution. The second integral may be readily computed if the pressure distribution of a given boat is known. 

A common approach to obtaining simplified analytical expressions of the Kelvin integral is to assume symmetric pressure distributions \citep[e.g.,][]{Havelock1909,Raphael1996} and Gaussian shapes \citep{Darmon2014,Brown1989,Pethiyagoda2017}. Gaussian pressure distributions provide a boat representation slightly more realistic than point distributions,  that has  the convenient property of self-similarity under the Fourier transform (the transform of a Gaussian is a Gaussian, e.g., equation \ref{eq: Gaussian}), 
\begin{equation}
p\left(x\right)=e^{-\pi^{2}F_{L}^{4}x^{2}},\;p\left(k,\theta\right)=\frac{1}{\pi F_{L}^{4}}e^{\left(-\frac{k^{2}}{4\pi^{2}F_{L}^{4}}\right)},\label{eq: Gaussian}
\end{equation}
which significantly simplifies the algebra.  Following \citet{Darmon2014},  \citet{Pethiyagoda2017,Pethiyagoda2018} used it to investigate the properties of the time-frequency representation of ship waves introduced by \citet{Sheremet2012}. 

The analytical boat model as defined by equation \ref{eq: Gaussian} is too crude for direct application to realistic boat traffic: the model is linear; the distribution is isotropic,  the ``draft'' is not independent of the ship size, and  the pressure distribution has infinite support (in other words ``Gaussian ships'' are infinite, have circles as depth contours, and large ships have shallow drafts; think of infinite ``fresbees").  The analytical wake model itself (equation \ref{eq: linear solution}) is linear and assumes constant depth.  For this discussion, however, as crude as it is, the Gaussian boat model introduces an additional characteristic spatial scale related to the boat, the length-based Froude number, $F_{L}=U/\sqrt{gL}$, can be used as a crude device to investigate qualitatively the effects of the ship size. For our purposes, the depth- and length-based Froude numbers may be regarded as a basic dimensions in a wake-parameter space (wake classification). 

The two scales represented by the Froude numbers play an essential role in determining the shape of the Gaussian-wake spectrogram, and thus answer the important question of how the instantaneous power of the wake is distributed in the time-frequency (see figure \ref{fig: Gaussian wakes}, which replicates figure 4 in \citet{Pethiyagoda2018}).  Subcritical wakes ($F_{H}<1$) are characterized by a two-valued dispersion curves:  a "chirp" branch, with  frequency increasing in time, corresponding to divergent waves, and a monochromatic (constant frequency) component corresponding to the transversal wave. As $F_{H}\nearrow1$, the frequency of the monochromatic tail of the wake decreases to zero. At values $F_{H}>1$ only the divergent wave remains: supercritical wakes have a single-valued branch of the dispersion curve, the chirp, corresponding to the divergent-wave (wedge-shaped) component of the wake. Figure \ref{fig: Gaussian wakes} shows that, for subcritical wakes, $F_{H}$ number controls the frequency of the monochromatic tail for $F_{H}<1$: higher $F_{H}$ means lower frequency, while the length-based Froude number $F_{L}$ controls the overall frequency of the wake, with low $F_{L}$ values (large ``ships'') corresponding low-frequency (long) wakes, and high $F_{L}$ (small ``ships'') resulting in high-frequency (short) wakes. Very low $F_{L}$ values produce wakes that are reminiscent of the drawdown waves generated by large ships, also referred to as depression or Bernoulli waves \citep{Bertram2014, Soomere2007}. These trends are useful in assessing basic capabilities of numerical Boussinesq models that have dispersion constraints, such as FUNWAVE-TVD.

\section{Field observations\label{sec: obs}}

\begin{figure}[h]
\begin{centering}
\textcolor{red}{\includegraphics[width=0.8\textwidth]{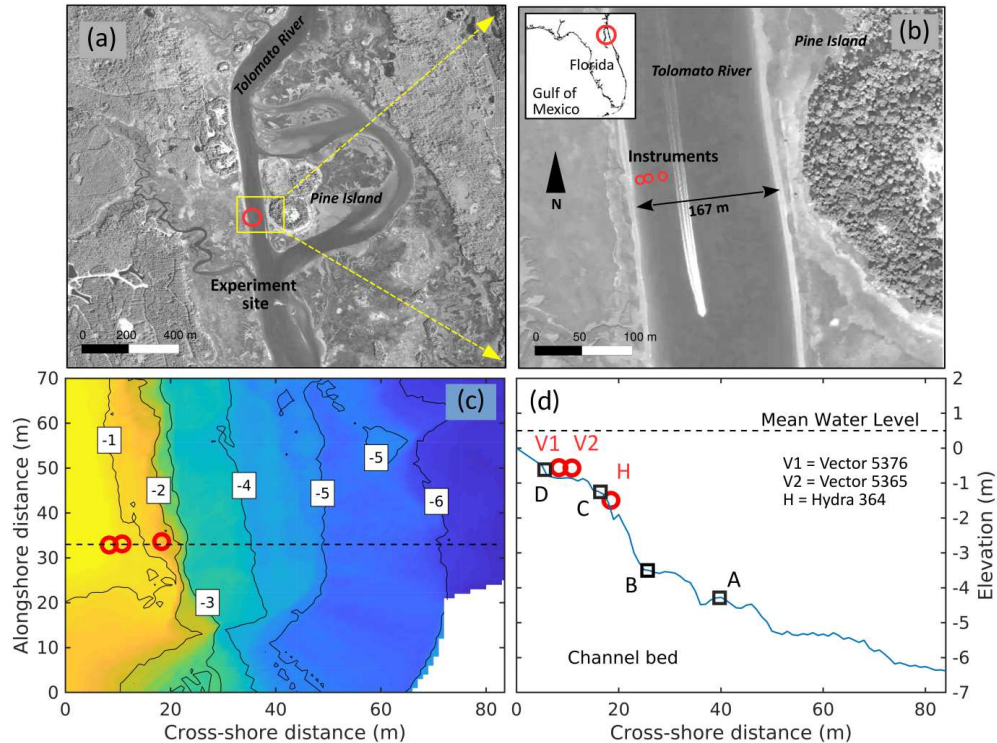}}
\par\end{centering}
\caption{Experimental site and instrument array configuration.  a) Experiment site along on the Intracoastal Waterway near Ponte Vedra Beach, Florida, USA. b) Detailed aerial view of the yellow rectangle in panel (a),  with the approximate location of the instruments. c) Bathymetry contours (in meters below mean water level).  d) Cross-channel profile with the location of the instruments and their position with respect to the mean water level. Red circles are used in all the panels to indicate the location of the instruments. Black squares mark the location of the output locations used in numerical simulations. }
\label{fig: expmap}
\end{figure}

Field observations were collected in 2017 from November 11th to December 4th along the Tolomato River (figure \ref{fig: expmap}) at a location inside the Guana-Tolomato-Matanzas National Estuarine Research Reserve (GTMNERR),  Ponte Vedra, Florida, USA.  The array of instruments deployed included two Nortek Vector ADV (Acoustic Doppler Velocimeter, numbers 5365 and 5376) and a Sontek Hydra ADV (364), which sampled  pressure, three-dimensional flow velocity, and acoustic backscatter at 8 Hz.  Fronting a channel with a width of approximately 150 m, the experiment site typically experiences very mild wave activity this period of the year, with rather infrequent weak westerly winds (<3 m/s; based on the GTMNERR Centralized Data Management Office):  over the duration of the entire experiment, wind waves (with mean period of approximately 2 s)  were observed only during a 3-hr interval on Nov. 24. Circulation is dominated by semi-diurnal tides with a range of approximately 1.5 m. 

\subsection{Boat traffic\label{subsec:traffic}}

\begin{figure}
\centering{}\includegraphics[width=0.8\textwidth]{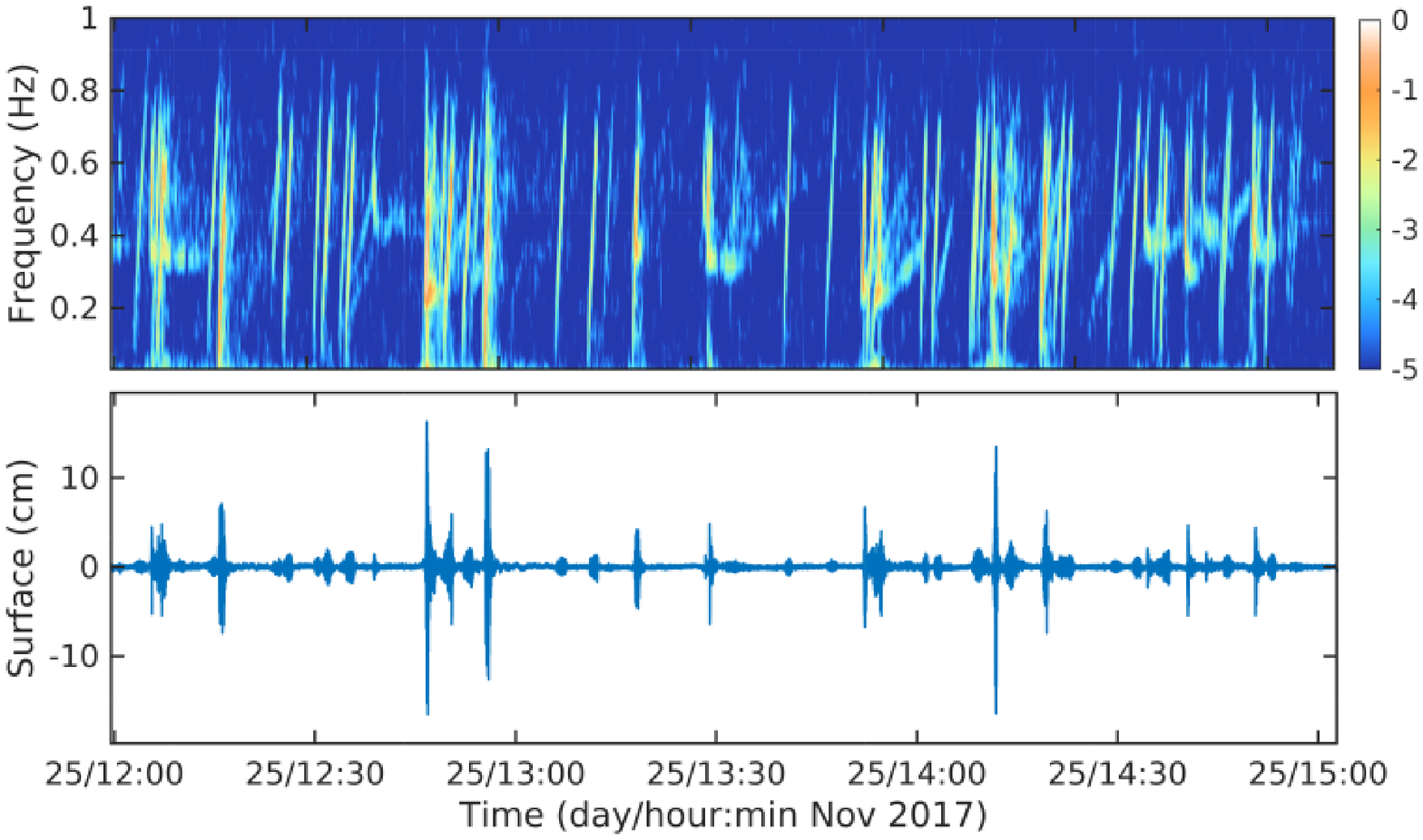}
\caption{An example of wake observations on Nov 25th, 2017. Top: spectrogram; bottom: pressure time series, collected at Hydra 364 station (see figure \ref{fig: expmap}) at  high tide, corresponding to local water depth of approximately 2 m). 
\label{fig: wakes}}
\end{figure}

By comparison to the natural, essentially wave-free conditions at the site at the time of the experiment, the wave generated by the rather intense boating during peak traffic hours was significant. In figure \ref{fig: wakes}, showing a time interval of 3 hours, one can count approximately 50 individual chirps (each chirp represents a wake, whether sub- or super-critical).  A crude, back-of-the-envelope" calculation, assuming  a characteristic wake duration of 1 minute and just "stringing" wakes one after another with no pause, suggests that every hour of peak boat traffic was roughly equivalent to 15 min of continuous, 1-min wave groups of 2-s waves of approximately 0.2 m average height.

\begin{figure}
\centering{}\textcolor{red}{\includegraphics[width=1\textwidth]{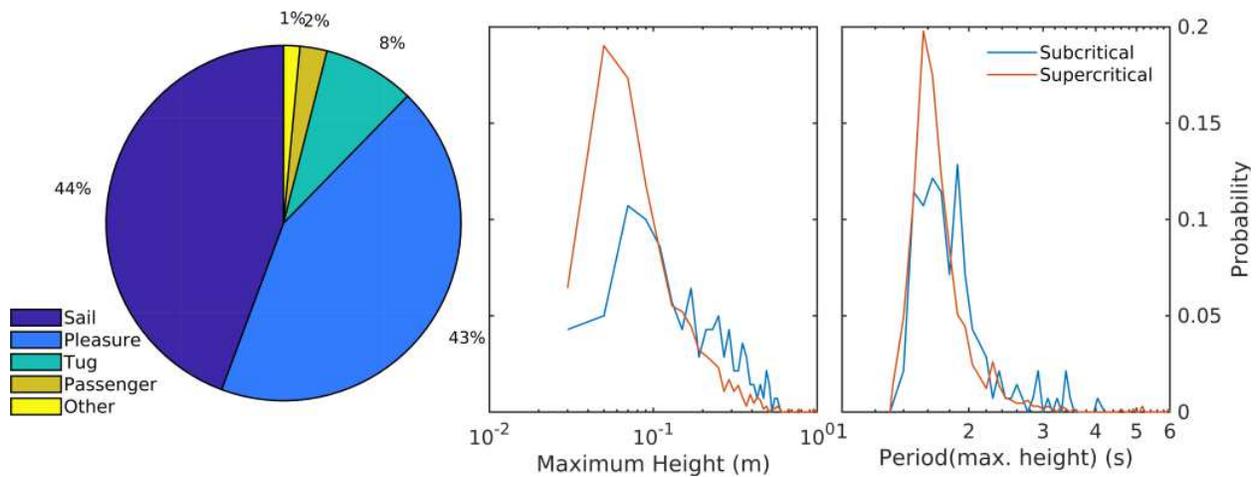}
\caption{Left: Chart of boat population as recorded in the AIS database between Nov. 17th to Dec. 4th, 2017, in a 1.2 km box centered on the location of the experiment site. Middle: probability density of trough-to-crest height for the largest wave in a wake group. Right: probability density of the period of the the largest wave in a wake group.}
\label{fig: traffic stat}}
\end{figure}

The variety of different vessel types present during the experiment is illustrated by the diversity of the spectrogram shapes recorded -- a summary of some statistical properties of the boat traffic is presented in figure \ref{fig: traffic stat}. During the experiment a total of 782 wakes were observed, with the vast majority supercritical, 140 sub-critical, identified based on the presence in the spectrogram of the characteristic monochormatic transversal wave (see, e.g., figure \ref{fig: Gaussian wakes}), and a few depression wakes. Based on the distribution of wakes recorded (figure \ref{fig: traffic stat}b), the largest waves in supercritical wakes were larger that  those in subcritical wakes, while the distribution of maximum-crest period was similar for the two types (figure \ref{fig: traffic stat}c). 

Searching through the U.S. Coast Guard Automatic Identification System (AIS) database for boats navigating the Tolomato River at the time of the experiment and in a  box of 1.2-km length centered on the experiment location retrieved only 203 records. The discrepancy between the number of observed wakes and AIS-recorded boats is probably due to a large population of small boats that are not equipped with AIS transmitter (ships over 300 tons and passenger vessels are required to have AIS transmitters, but recreational vessels may opt out of this system; \citealp{WestMarineAIS}). Although representing only a small section of the traffic recorded at the site, the AIS records do provide some indication of the major traffic constituents  (figure \ref{fig: traffic stat}a): sailing and pleasure boats accounted for 88\% of the population, with tug, towing, passenger, and other unidentified boats comprising the remaining 12\% (figure \ref{fig: traffic stat}a).

We suggested above (section \ref{subsec: climate}) that the practical impossibility  to provide accurate initial conditions for boat traffic simulations is fundamental to the statistical nature of the wake climate problem. To investigate this idea, we attempted to associate specific wake observations with AIS recorded boats, but despite our best efforts, we were unable to do so. It is important to note here that we do not consider this a failure of the AIS database\footnote{The AIS is an important tool whose purpose is to monitor vessel traffic and characteristics (length, width, draft and speed) in busy commercial shipping channels, where the majority of crafts are large container/tanker-type vessels. For this function the AIS database is no doubt accurate and practical.}.  Rather, we interpret our lack of success to reflect the fundamental difficulty of trying to match in situ observed wakes with individual boats recorded by a monitoring system.  Even if the boat has a transponder and its trajectory is recorded,  at the AIS message rate of 1 every 30 s  a boat moving at 5 m/s, subcritical in 10-m depth, will travel 150 m between messages. The position uncertainty  increases significantly if one allowing for speed variations (slow-speed zones or boat encounters),  for the time delay between wake generation and observation,  and for possible overlaps between wakes created by boats navigating close to each other in busy traffic. More generally, and returning to the point of ill defined forcing, our lack of success illustrates how difficult it is to reliably associate a wake observed in situ with a boat recorded by a traffic monitoring system based on  (essentially sparse) sampling of the Lagrangian traffic flow.  

\subsection{A characterization of observed wakes} 

\subsubsection{Linear characteristics}

\begin{figure}[h]
\begin{centering}
\includegraphics[width=0.8\textwidth]{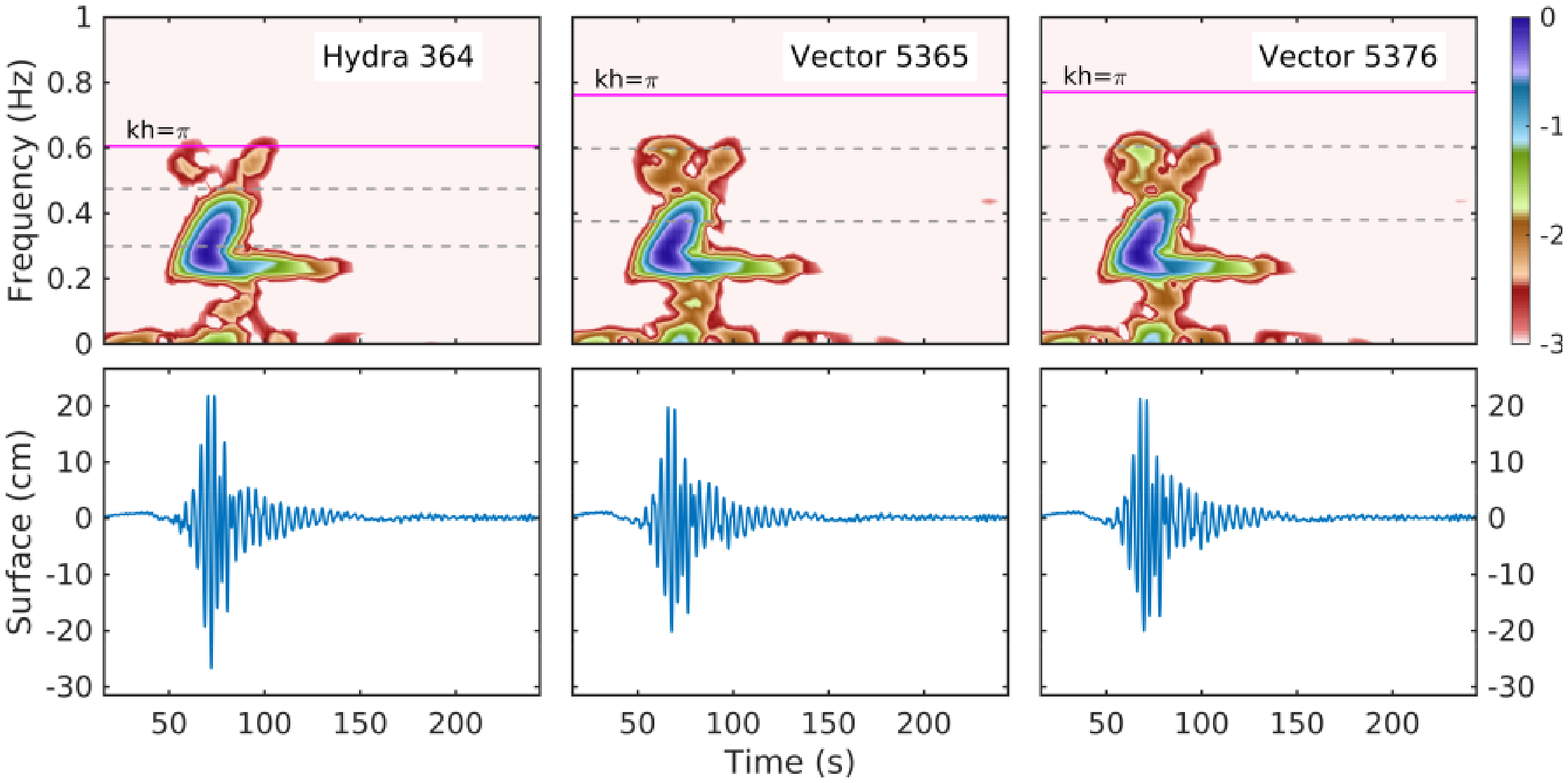} \vspace*{\medskipamount}
\par\end{centering}
\begin{centering}
\begin{tabular}{|c|c|c|c|c|c|}
\hline 
 &  & \multicolumn{2}{c|}{{\small{}Main wake}} & \multicolumn{2}{c|}{{\small{}Harmonic}}\tabularnewline
\hline 
{\small{}Instrument} & {\small{}Depth (m)} & {\small{}$f$ (Hz)} & {\small{}$kh$} & {\small{}$f$ (Hz)} & {\small{}$kh$}\tabularnewline
\hline 
\hline 
{\small{}Hydra 364} & {\small{}2.12} & {\small{}0.34} & {\small{}1.18} & {\small{}0.34} & {\small{}2.70}\tabularnewline
\hline 
{\small{}Vector 5365} & {\small{}1.34} & {\small{}0.34} & {\small{}0.88} & {\small{}0.34} & {\small{}1.79}\tabularnewline
\hline 
{\small{}Vector 5376} & {\small{}1.31} & {\small{}0.34} & {\small{}0.86} & {\small{}0.34} & {\small{}1.75}\tabularnewline
\hline 
\end{tabular}
\par\end{centering}
\caption{Example of a subcritical boat wake created by a slow-moving vessel traveling southward, recorded at approximately 13:00 hr local time on Nov. 22nd, 2017. Top row: spectrogram normalized to unit maximum value, $\log_{10}$ scale; bottom row: free surface elevation. Magenta line marks the frequency corresponding to the weak-dispersion limiting value $kh=\pi$ at the isobath of the sensor. Gray lines correspond to  $kh=1,2$. Table: estimates of the mean frequency and $kh$ for the main wake body and harmonic.
\label{fig: sub struct}}
\end{figure}

\begin{figure}[h]
\begin{centering}
\includegraphics[width=0.8\textwidth]{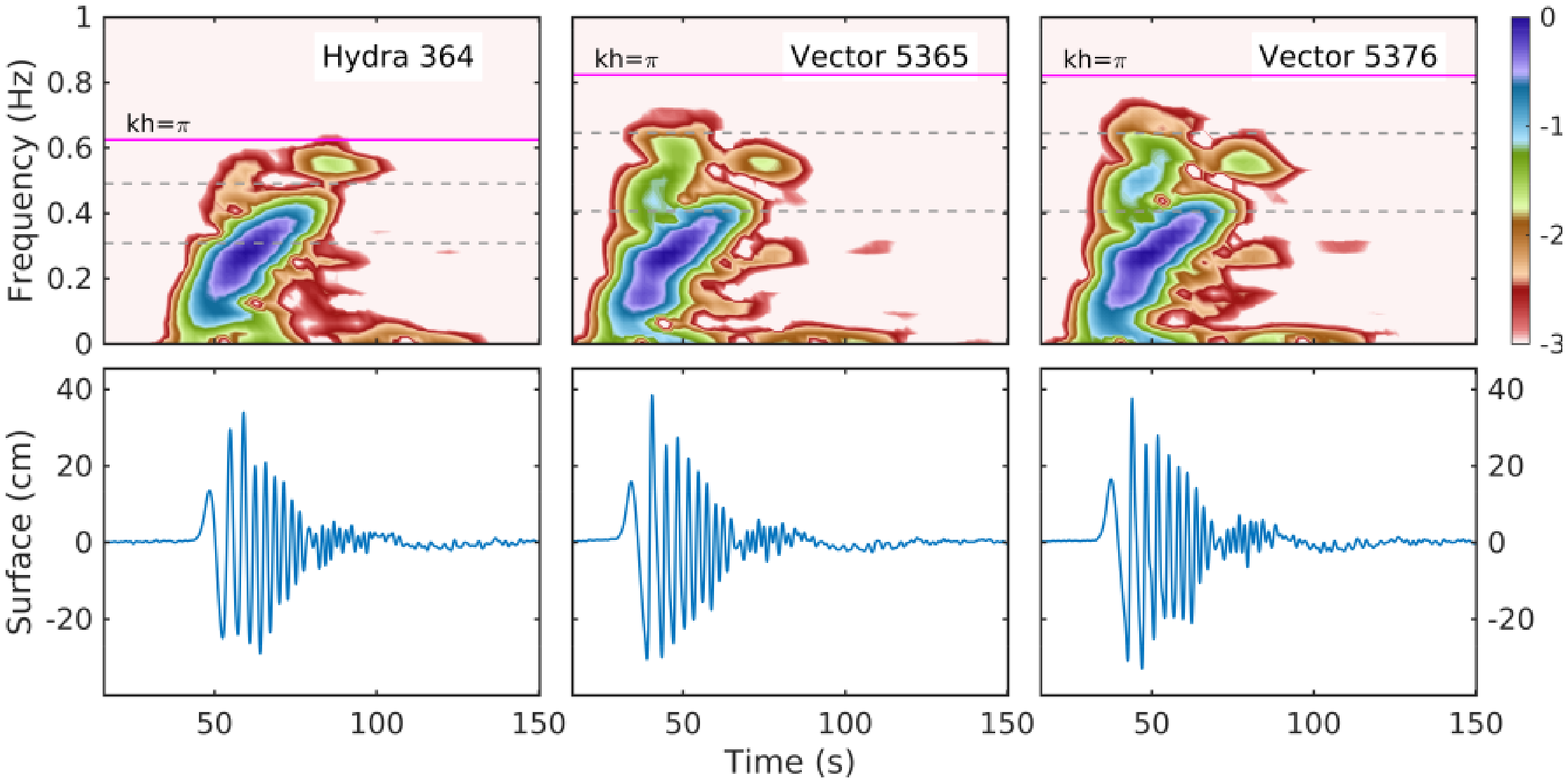} \vspace*{\medskipamount}
\par\end{centering}
\begin{centering}
\begin{tabular}{|c|c|c|c|c|c|}
\hline 
 &  & \multicolumn{2}{c|}{{\small{}Main wake}} & \multicolumn{2}{c|}{{\small{}Harmonic}}\tabularnewline
\hline 
{\small{}Instrument} & {\small{}Depth (m)} & {\small{}$f$ (Hz)} & {\small{}$kh$} & {\small{}$f$ (Hz)} & {\small{}$kh$}\tabularnewline
\hline 
\hline 
{\small{}Hydra 364} & 1.98 & {\small{}0.28} & {\small{}0.88} & {\small{}0.53} & {\small{}2.28}\tabularnewline
\hline 
{\small{}Vector 5365} & {\small{}1.14} & {\small{}0.28} & 0.64 & {\small{}0.53} & {\small{}1.44}\tabularnewline
\hline 
{\small{}Vector 5376} & {\small{}1.14} & {\small{}0.28} & 0.64 & {\small{}0.53} & {\small{}1.44}\tabularnewline
\hline 
\end{tabular}
\par\end{centering}
\caption{Example of a supercritical wake recorded at approximately 13:45 hr,  on Dec. 4th, 2017. This is one of the largest wakes recorded during the experiment. Top row: normalized spectrogram, $\log_{10}$ scale, maximum value 1; bottom row: free surface elevation.  Magenta line marks the frequency corresponding to the weak-dispersion limiting value $kh=\pi$ at the isobath of the sensor. Gray lines correspond to  $kh=1,2$. Table: estimates of the mean frequency and $kh$ for the main wake body and harmonic.\label{fig: sup struct}}
\end{figure}

\begin{figure}
\begin{centering}
\includegraphics[width=0.8\textwidth]{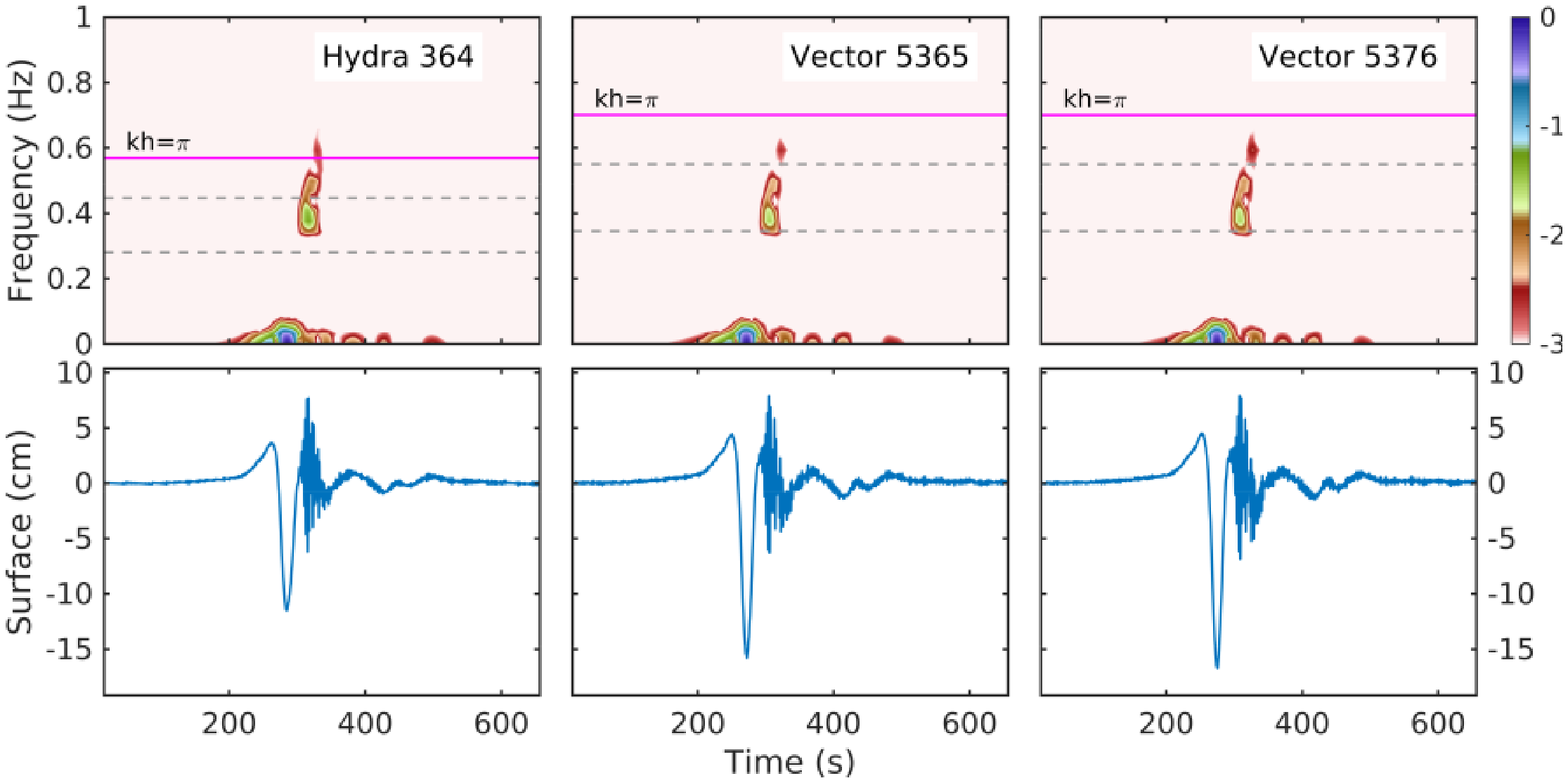}\vspace*{\medskipamount}
\par\end{centering}
\begin{centering}
\begin{tabular}{|c|c|c|c|}
\hline 
 &  & \multicolumn{2}{c|}{{\small{}Trailing wave}}\tabularnewline
\hline 
{\small{}Instrument} & {\small{}Depth (m)} & {\small{}$f$ (Hz)} & {\small{}$kh$}\tabularnewline
\hline 
\hline 
{\small{}Hydra 364} & 2.4 & {\small{}0.42} & {\small{}1.8}\tabularnewline
\hline 
{\small{}Vector 5365} & {\small{}1.43} & {\small{}0.42} & {\small{}1.31}\tabularnewline
\hline 
{\small{}Vector 5376} & {\small{}1.44} & {\small{}0.42} & {\small{}1.31}\tabularnewline
\hline 
\end{tabular}
\par\end{centering}
\caption{Example of a depression wake created by a large ship traveling northward, recorded at approx. 7:20 hr local time on Dec. 1, 2017. Although we could not identify the ship that generated this wake, a likely candidate could a passenger ship with a length in the order of 50 m, width of 15 m and draft of 2 m.Top row: normalized spectrogram, log scale, maximum value 1; bottom row: free surface elevation.  Magenta line marks the frequency corresponding to the weak-dispersion limiting value $kh=\pi$ at the isobath of the sensor. Gray lines correspond to  $kh=1,2$. Table: estimates of the mean frequency and $kh$ for the high-frequency wave typically trailing the drawdown wake.
\label{fig:  down struct}}
\end{figure}

Despite its simplicity, the analytic linear model provided by equation \ref{eq: linear solution} appears to be useful in classifying wake observations in the field. Figures \ref{fig: sub struct}-\ref{fig: sup struct} illustrate easily-recognizable field realizations of three basic wake types suggested by the analytical model: a subcritical wake characterized by a combination of divergent and transversal components, displaying a characteristic ``L'' shape, figure \ref{fig: sub struct}; a supercritical wakes showing a single chirp (divergent wave), and a depression wake, figure \ref{fig: sup struct}. Supercritical wakes were dominate the population of waves observed (640 out of 782) and have higher maximum amplitude (figure \ref{fig: traffic stat}). Figure \ref{fig:  down struct} shows the largest supercritical wake, recorded at 14:00 hrs on Dec. 4th, 2017. After correcting pressure records for surface elevation (see section \ref{sec: methods}),  its highest wave is approximately 60 cm trough-to crest at the shallowest sensor.  Only two depression (drawdown) wakes were recorded by the experiment, e.g. figure \ref{fig:  down struct}. Although the type of the vessel has not been identified in the AIS database, the relatively low maximum height of the wake and the trailing monochromatic tail suggests a slow moving vessel, possibly a sail boat, navigating southward along the Tolomato River.  

\subsubsection{Nonlinear shoaling effects}

While spectrogram shapes estimated based on measurements at the most offshore sensor (Hydra 364, figure \ref{fig: expmap}a) match the solutions of the linear model  given by equation \ref{eq: linear solution}), at shallower sensors they deviate systematically from the linear solution. The spectrogram develops additional peaks that increase in power shoreward, (quite obvious in figure \ref{fig: sup struct}) located at time-frequency positions consistent to instantaneous harmonic frequencies of the linear solution. These  peaks represent frequency component excited during shoaling and are fundamentally effects of the nonlinear shoaling evolution of the wake (linear evolution preserves the number of waves, and consequently, the spectral shape).  
\begin{figure}[h]
\begin{centering}
\includegraphics[width=0.8\textwidth]{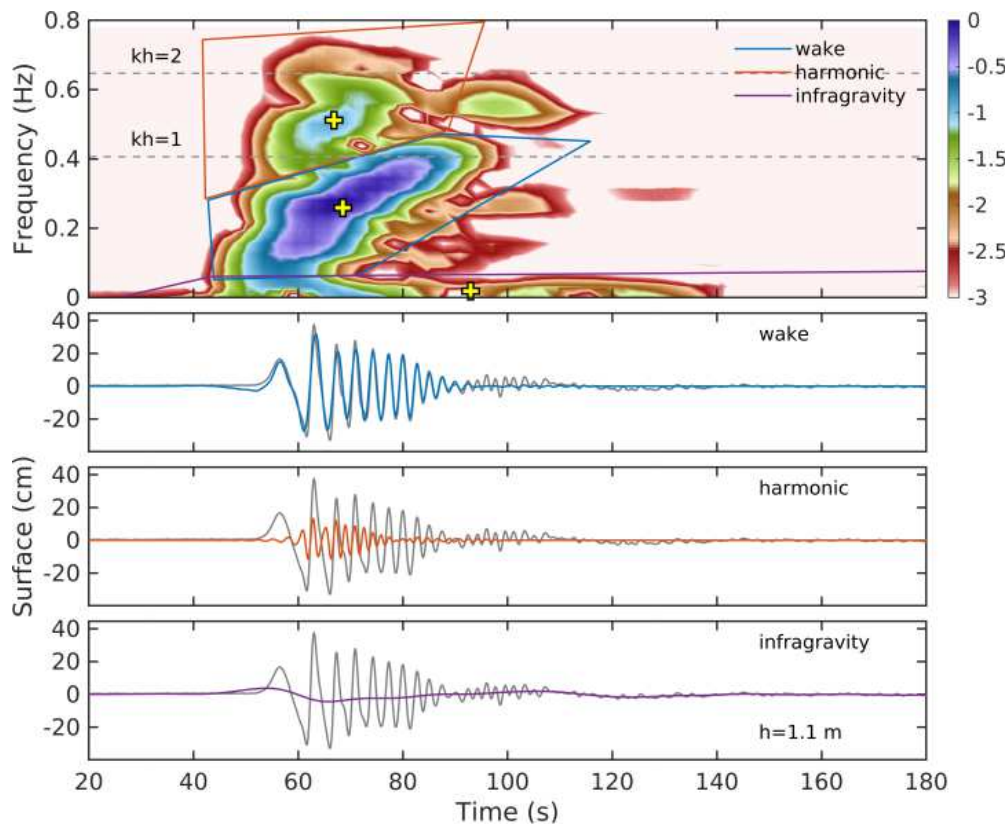}
\par\end{centering}
\caption{Decomposition of the supercritical wake recorded at 13:45 hr,  on Dec. 4th, 2017 into components identified by their time-frequency domain. The time-frequency filter is applied to measurements of free-surface elevation measurements collected by  Vector 5365 (1.1-m water depth, see figure \ref{fig: expmap}). Top panel: Spectrogram with the time-frequency domains corresponding to different wake components identified by polygons (blue – primary wake chirp; red – harmonic of the primary chirp; purple – infragravity wave). Yellow crosses correspond to the center of mass of wake components. Bottom three panels: Filtered time series corresponding to the components identified in the top panel: wake, harmonic, and infragravity. The full pressure signal is plotted (gray) as reference. \label{fig: sup comps}}
\end{figure}

Using the time-frequency filter (section \ref{sec: methods}), one can reconstruct separately the time series corresponding to each of the nonlinear components (new spectrogram peaks) and examine their role in reassembling the full observed free surface elevation (figure \ref{fig: sup comps}).  The primary chirp of the wake (blue) is has a general sinusoidal form (i.e., the frequency and amplitude oscillation do not deform the shape significantly over an instantaneous period, figure \label{fig: sup comps}). The secondary chirp that parallels the primary one and is positioned approximately along the double frequency of the primary chirp has two readily detectable contributions: in general, its crests work toward amplifying the crests and flattening the trough, as in a regular Stokes wave \citep[e.g.,][]{Whitham1974}, but toward the front of the wake the harmonic crests shift forward, increasing the steepness of the front of the wake. The long-wave component (purple) appears to follow the typical description of the long wave bound to the wake group \citep[e.g.,][]{Longuet-Higgins1962}.

\begin{figure}[h]
\begin{centering}
\includegraphics[width=1\textwidth]{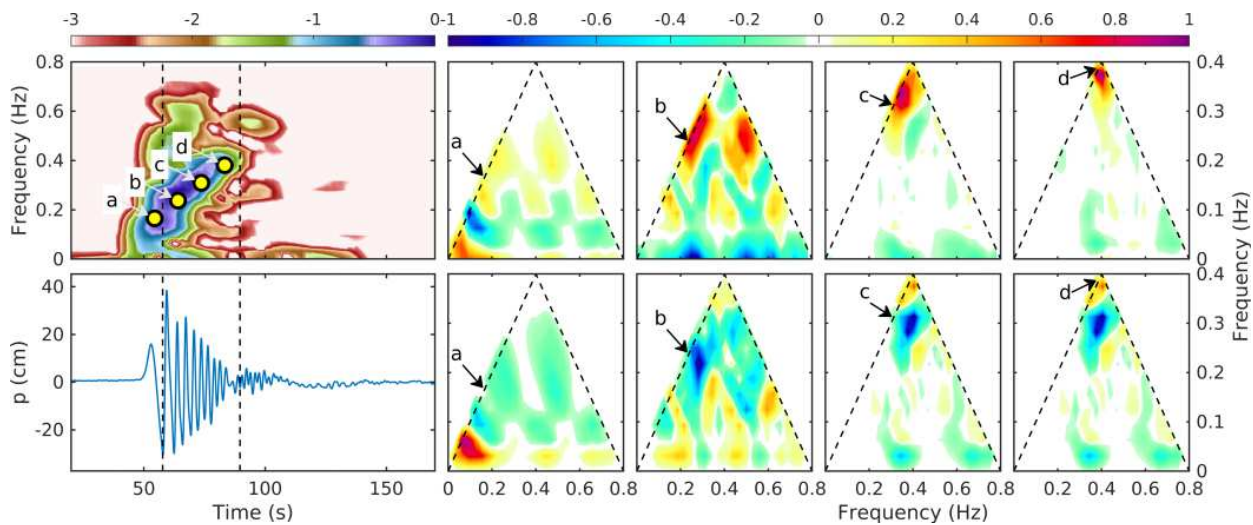}
\par\end{centering}
\caption{Biperiodogram of the pressure signal of the Dec. 4, 2017 supercritical wake, recorded by Vector 5365 (1.1-m water depth, see figure \ref{fig: expmap}). Leftmost column: spectrogram and time series. Rest of the panels, top: real part of the biperiodogram, $\Re\left\{ B\right\} $, a measure of the ``local'' skewness. Bottom: The imaginary part of the periodogram, $\Im\left\{ B\right\} $, a measure of the ``local'' asymmetry of the wake. The  center of the time windows corresponding to each periodogram is marked by dots in the spectrogram panel, located at the estimated instantaneous frequency of the primary chirp (maximum amplitude). The time-length of the window used is shown by dashed red lines. In the biperiodogram panels, arrows indicate the point $(f_{1},f_{1})$, where $f_{1}$ is instantaneous frequency of the primary chirp at the center of the time window.\label{fig: bsptg}}
\end{figure}

The biperiodogram phase-coupling analysis (figure \ref{fig: bsptg}) of the free surface elevation recorded for this wake at Vector 5365 supports this interpretation. As discussed in section \ref{sec: methods}, the structure and interpretation of the biperiodograms  is similar to the interpretation of the bispectrum information: the distribution in the plane $(f_{1},f_{2})$ indicates a possible phase relation between the frequencies $f_{1}$, $f_{2}$ and $f_{1}+f_{2}$. Although the  Fourier window covers a significant time span of the wake (dashed lines in the left-most panels in figure \ref{fig: bsptg}), the variability of the biperiodogram seems sufficient to identify and follow consistent phase relations as they evolve (as the window slides) over the wake time span. 

The strength of the local phase-coupling evolves along the wake duration. It starts as weak and nearly-linear at the front of the wake (window centered at point "a" left-most panels in figure \ref{fig: bsptg}), and strengthens considerably afterwards. The biperiodograms corresponding to  windows centered at points "b"  to "d" exhibit strong  coupling between the principal  wake and its harmonic (red blobs on the first diagonal of the real part of the biperiodogram, figure \ref{fig: bsptg}, top row),  indicating that the harmonic of the primary wake contributes to skewing positively the time series, i.e., wave peaking. The imaginary part of biperiodogram (figure \ref{fig: bsptg}, bottom row) shows negative peaks, slightly off the first diagonal, showing that the harmonic (or a band close to it) contributes to negative asymmetry of the time series, i.e., steepening of wave fronts.  This is consistent with the behavior described by the time-frequency filtering method. 
 
While the nonlinear shoaling evolution seen in figure  \ref{fig: bsptg} is somewhat similar to the Stokes-type evolution described by \citet{Pethiyagoda2017}, it differs in several important respects: 1) it occurs over relative short distances; 2) it generates strong \emph{bound} harmonics that do not satisfy the dispersion relation, and whose role is to distort the shape of the initially non-skewed, symmetric wake by crest peaking and wave-front steepening; and 3) it also generates a low-frequency wave with periods > 10 s. This type of evolution is more akin to  nonlinear shoaling effects observed in (stationary) ocean waves \citep[e.g.,][and many others]{Elgar1985,Sheremet2003,Sheremet2016T}. In the case of ocean waves, the nonlinear shoaling process is generally described as a nonlinear phase-coupling and energy transfer between the peak of the spectrum (typically swell) and higher/lower frequencies. Because the higher frequencies are normally in a non-resonant regime and have initially very low amplitudes, they develop as bound waves, whose presence is expressed through the deformation of the swell shape from almost sinusoidal to a nearly saw-tooth form: shoaling swells  peak, with the wave front  steepening while the back flattens into a wide trough. Lower frequencies generated through nonlinear shoaling are usually called infragravity waves, a name that is meant to express the fact that (under local normal conditions) they   cannot be generated directly by wind. As is the case of ocean waves, the shoaling and breaking processes are expected to have profound consequences for sediment transport \citep[e.g.,][and others]{Nielsen2006,Ruessink2009,VanderA2010}.

\section{Numerical Simulations\label{sec: num}}

The analytic solution (e.g., figure  \ref{fig: Gaussian wakes}) and field observations suggest a few simple questions: How much of the parameter space can the model cover with reasonable accuracy? How well does the model describe the processes of nonlinear shoaling of wakes in this parameter space? What are the weaknesses of the model and how do they affect its general performance? 

The Boussinesq FUNWAVE-TVD implementation was selected here because it fits the shallow urbanized coastal environments, and is mature and well understood  \citep[e.g.,][]{Tehranirad2011, Shi2012, Malej2015, Shi2016,Malej2019}. The model's ability to simulate wakes produced by large commercial ships has recently also been tested (Malej, personal communication).  This helps with the validation efforts, but it does not eliminate a basic  challenge:  inaccurate information about offshore forcing (boat traffic, see (section \ref{subsec: traffic})) means that it is uncertain whether the wave field to be simulated meets the weak-dispersion condition.  Large ships  moving slowly likely generate long waves, but over the entire population of boats comprising the traffic conditions may vary substantially.  
The analytic model (equations \ref{eq: linear solution} and \ref{eq: Gaussian} and  figure \ref{fig: Gaussian wakes} ) provide some useful guidance, but the model itself is too crude to be applicable to observations. To circumvent problem we take here two  steps: 1.  we test the numerical model on the analytical solution, and  2. using FUNWAVE-TVD's own boat representation modules, we select a set of boat/navigation parameters that fit the description of three representative wakes, and examine its performance in reproducing observed features of nonlinear shoaling evolution of the wakes.    
\subsection{Analytical linear Gaussian wakes}
\begin{figure}[H]
\begin{centering}
\includegraphics[width=0.7\textwidth]{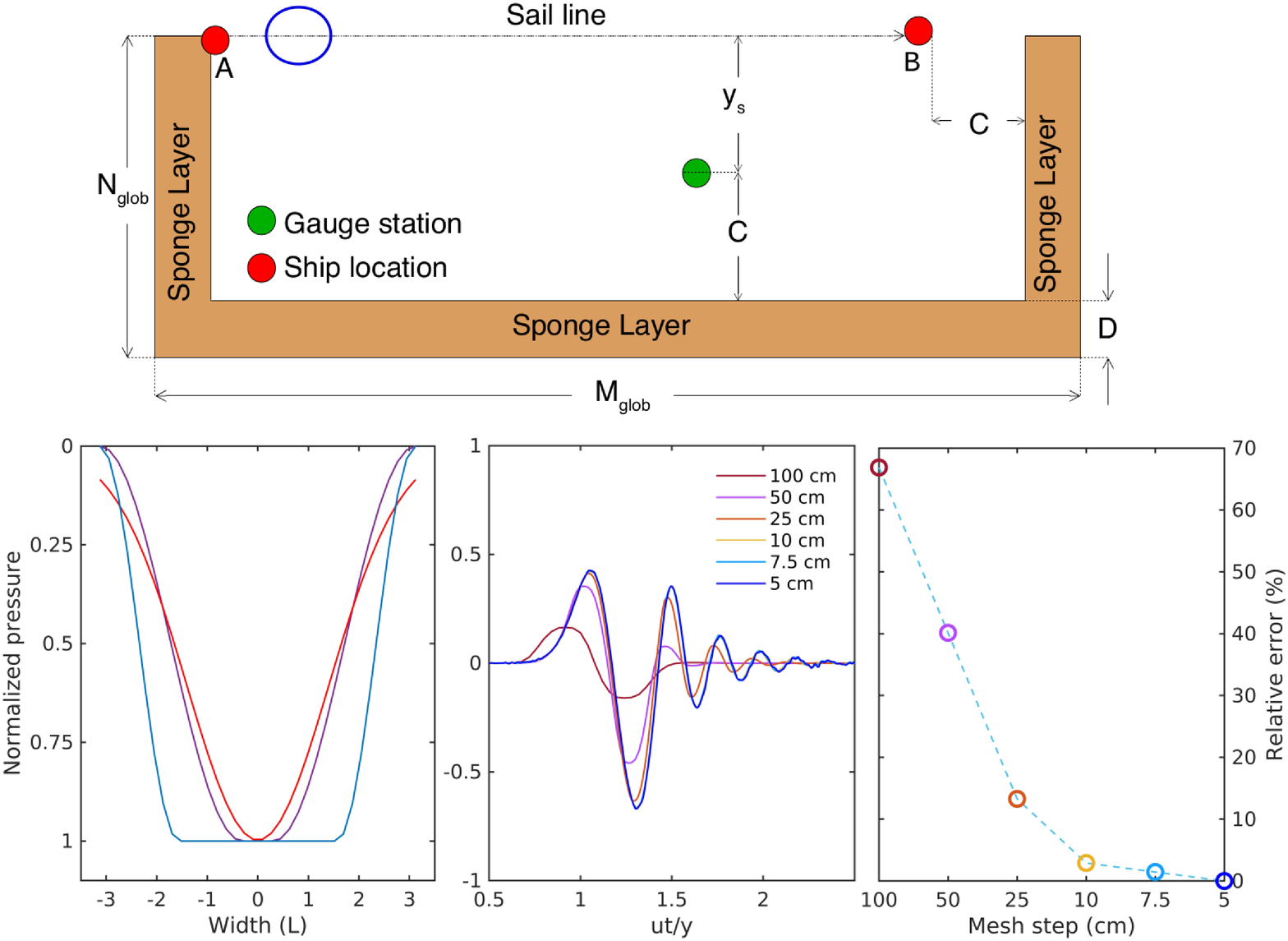}\ \ 
\par\end{centering}

\begin{centering}
\bigskip{}
\begin{tabular}{|ll|ll|ll|}
\hline 
{\small{} Parameter}     & {\small{} Value} & {\small{} Parameter} & {\small{} Value} & {\small{} Parameter} & {\small{} Value}\tabularnewline
\hline 
{\small{} $F_{H}$}        & {\small{}1.5 }        & {\small{} Grid step}           & {\small{} 0.075} m      & {\small{} Ship speed} & {\small{} 3.13 m/s}\tabularnewline
{\small{} $F_{L}$}        & {\small{} 0.4}        & {\small{} $y_{s}$, $C$}      & {\small{} 25 m, 80 m} & {\small{} Ship length} & {\small{} 2.038 m}\tabularnewline
{\small{} Depth}          & {\small{} 0.44 m} & {\small{} Sponge width}   & {\small{} 2 m}              & {\small{} Draft}             & {\small{} 0.01 m}\tabularnewline
{\small{} $M_{glob}$}& {\small{} 4453}    & {\small{} Model time $t$} & {\small{} 79.87 s}     & {\small{} $A$}         & {\small{} 2.0 m}\tabularnewline
{\small{} $N_{glob}$} & {\small{} 1426}    & {\small{} Pressure}              & {\small{} Gaussian} & {\small{} $B$}            & {\small{} 254 m}\tabularnewline
\hline 
\end{tabular}
\par\end{centering}
\caption{Model setup and convergence tests for comparisons with analytical Gaussian wakes.  Left:  Pressure distribution used in the simulations: Gaussian distribution (red); FUNWAVE-TVD PST-I pressure function ($\alpha_{1}=\alpha_{2}=\beta=0.5$ -- blue; and $\alpha_{1}=\alpha_{2}=\beta=0.1$ -- purple). Center: Pressure  time series and relative errors for the mesh sizes used in convergence tests (mesh sizes: 100, 50, 25, 10, 7.5, and 5 cm, corresponding to  10, 20, 22, 40, 65 and 105 grid points/shortest wavelength. Table: Numerical simulation parameters for the tests with analytical Gaussian wakes.
\label{fig: Gaussian ship}}
\end{figure}

Although the analytical  Gaussian model  does not include nonlinear effects and defines unrealistic boat shapes, it is still useful as a benchmark because the Froude-number classification it defines is relevant for field observations.  Gaussian perturbations were simulated using the type-1 pressure distribution provided in FUNWAVE-TVD as a ship-wake module \citep[e.g.,][]{Ertekin1986, Wu1987, Torsvik2009}. The module defines a rectangular horizontal footprint  $x-x_{c}(t)\in\left[-\frac{L_{1}}{2},\frac{L_{1}}{2}\right]\times\left[-\frac{L_{2}}{2},\frac{L_{2}}{2}\right],$ where $x=(x_{1},x_{2})$ is the vector of position in the horizontal plane, $x_{c}(t)$ is the position of the center of the ship, and $L_{1,2}$ are the lengths of the rectangle sides. The vertical profile of the ship is rounded rectangle, with ``roundness'' parameters $0\le\alpha_{1,2},\beta\le1$. Gaussian pressure distributions  (equation \ref{eq: Gaussian}) may be approximated by setting $L_{1}=L_{2}$ and $\alpha_{1,2}=\beta=0.01$.
 Numerical convergence tests with decreasing mesh sizes (figure \ref{fig: Gaussian ship}) simulated a Gaussian pressure distribution moving at a speed 3.13 m/s on a  334 m x 107 m domain bounded by a 2-m sponge layer (figure \ref{fig: Gaussian ship}.). This particular geometry was chosen so as to return after re-scaling a length unit of 1 m.  Based on these tests, a mesh size  of 7.5 cm was selected as optimal for the numerical experiments.

\begin{figure}
\centering{}\includegraphics[width=0.8\textwidth]{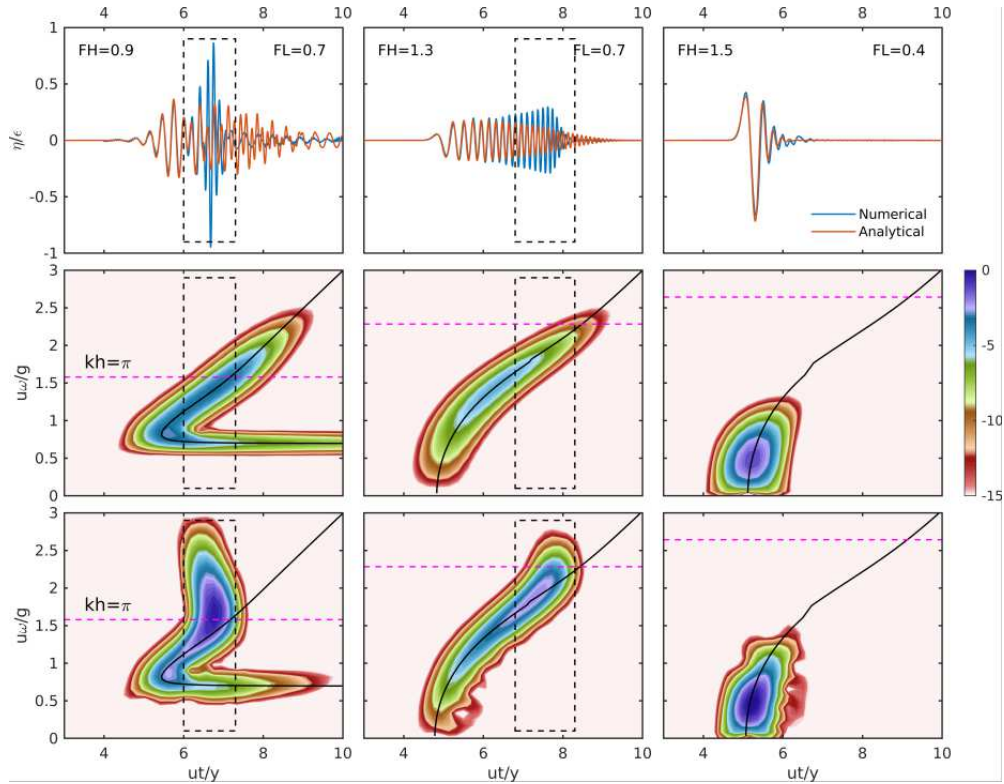}
\caption{Comparison of the analytical solution (equation \ref{eq: linear solution}) for a Gaussian wake and FUNWAVE-TVD simulations. Left: subcritical wake. Middle: supercritical wake. Right: ``depression'' wake. Top row: free surface elevation. Middle row: analytical solution. Bottom row: numerical simulation. The dashed black rectangles mark the approximate time interval in which the main chirp extends outside the domain $kh<\pi $. The dashed magenta line marks the normalized frequency corresponding to  $kh=\pi $ (values change in different columns because the frequency is normalized by the boat speed).   \label{fig: Gaussian numerical}}
\end{figure}

Figure  \ref{fig: Gaussian numerical} summarizes the performance of  FUNWAVE-TVD for three representative wakes, defined in the the two dimensional parameter space of the two Froude numbers, as:  $(F_H,F_L)=(0.9,0.7)$, subcritical; $(F_H,F_L)=(1.3,0.7)$, supercritical; and $(F_H,F_L)=(1.5,0.4)$, depression.  Both the sub- and supercritical  wakes spectrograms extend at some time intervals past the weak-dispersion validity limit of $kh\approx \pi $. In the $kh> \pi $ domain  the numerical solution preserves for a while the correct phase, but time-domain amplitudes are amplified, which causes both an accumulation of power density and the appearance of spurious high-frequency components. As a result, nonlinearity presumably becomes active in FUNWAVE-TVD and possibly leads to breaking. This dampens completely the chirp component, which disappears completely after reaching the $kh\approx\pi$ domain. 

The ``depression'' wake chosen for testing is well described due to its very low center of mass in frequency. This suggests that  FUNWAVE-TVD is well suited for modeling large container-type vessels, where the signature wave shape is indeed the depression (or drawdown) wave.

\begin{figure}
\begin{centering}
\includegraphics[width=0.7\textwidth]{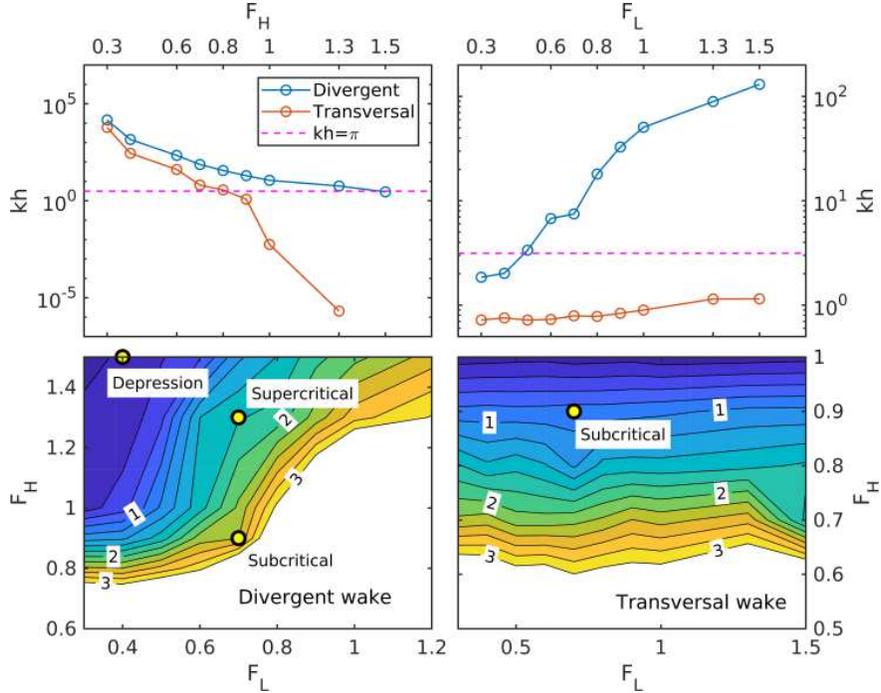}
\par\end{centering}
\caption{The distribution of $kh$ values in the Gaussian-wake parameter space of $ (F_H,F_L)$ defined by the analytical model (equation \ref{eq: linear solution}). Top left: distribution of mean  $kh$ values as  function of  the depth-based Froude number
$F_{H}$  for $F_{L}=0.7$. Top right:   distribution of mean  $kh$ values as  function of  the length-based Froude number $F_{L}$ for  $F_{H}=0.9$). Mean $ kh$ values correspond to the centers of mass of the divergent (chirp) and transversal  (monochromatic) wake components.  Bottom left mean $kh$ for the divergent wake as a function $F_{H}$ and $F_{L}$. Bottom right: mean $kh$ for the transversal wake as a function $F_{H}$ and $F_{L}$ (the monochromatic transversal wave exists only for $F_{H}<1$). The contoured regions are correspond to valid numerical representations ($ kh<\pi $). FUNWAVE-TVD yields invalid results in the regions left blank. Yellow dots correspond to the wakes shown in figure \ref{fig: Gaussian numerical}).
\label{fig: Gaussian space}}
\end{figure}

The performance of FUNWAVE-TVD  over a a wider population of Gaussian linear wakes is shown in figure \ref{fig: Gaussian space}.   The domain of validity of the FUNWAVE-TVD models is mapped onto the Froude-number space by associating to each wake  $(F_H,F_L)$  characteristic values of $kh$ corresponding to the center of mass of the divergent (chirp) and the transversal (monochromatic) components of the wake (the latter exists only for subcritical wakes). As expected,  wakes with large $F_H$ and low $ F_L$  values (boats large enough and moving fast enough; e.g., lower-left corner of the wake table in figure \ref{fig: Gaussian wakes}) ) tend to satisfy the weak-dispersion constraint.  As $ F_L $ decreases (smaller boats), the divergent wave (chirp wake) becomes the most problematic --   Gaussian-test maps suggest a rather severe $kh\approx 0.8$ lower boundary for the validity of FUNWAVE-TVD representation. Because transversal waves (monochromatic wake) is typically low frequency, the FUNWAVE-TVD representation appears to be  uniformly valid
for $F_{L}$ for any fixed value of $F_{H}\apprge0.7$. 

The tests on analytical Gaussian wakes illustrating the distortion introduced in the model representation by strongly-dispersive wake components, and help to qualitatively circumscribe the domain of applicability of FUNWAVE-TVD in the $(F_{H},F_{L})$ space. Based on these tests, one may conclude that FUNWAVE-TVD describes correctly the linear evolution of the wake provided that the weak-dispersion constraint is met. 

\subsection{More realistic wakes: nonlinear evolution}

The discussion of  Gaussian wakes provided a simple example of a parameter space for a wake population, but because its  limitations, is not useful for practical applications. An important element not represented in these tests is the nonlinear shoaling evolution of the wakes, of paramount importance for sediment transport. 

At the time of this study, our ability to associate observed  wakes with their generating  boats was at best ambiguous. Without such information, detailed phase-resolving comparisons between observations and numerical simulations \citep[e.g.,][]{David2017} cannot be conducted, because the source of possible errors cannot be assessed.

\begin{figure}
\begin{centering}
\includegraphics[width=0.6\textwidth]{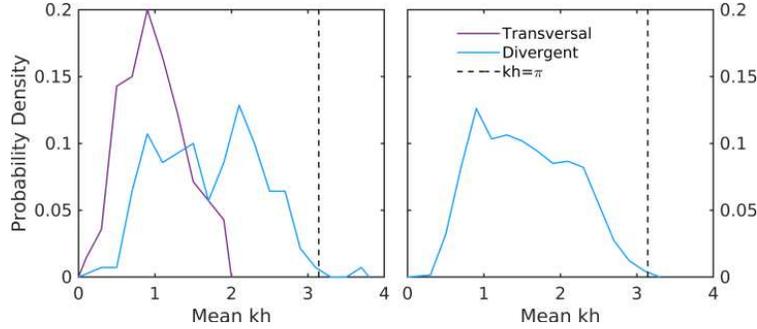}
\par\end{centering}
\caption{Estimates of probability density of mean $kh$ values for wake subcritical
(left) and supercritical (right) wakes, based on the data collected
at the location of Vector 5365 (see figure \ref{fig: expmap}). \label{fig: kh stats}}
\end{figure}
 
However, the  excellent performance demonstrated by linear Gaussian-wake tests for $kh<\pi$ carry over well to field observations \citep[e.g.,][]{David2017}). FUNWAVE-TVD is a well established model for simulating  nonlinear wave evolution in general, including wave-current interactions \citep{Shi2012} and nonlinear \citep{Choi2018}. We propose here that, relying on the large amount of knowledge about the general performance of Boussinesq models accumulated through previous studies, some insights can still be gained by examining the performance of the model itself.  Assuming that  can reasonably conclude that FUNWAVE-TVD describes well the linear and nonlinear weakly-dispersive wave fields processes in general, we can focus on an important remaining question: how well is the nonlinear shoaling process characterized for an entire  {\em population} of wakes. 

\begin{figure}
\begin{centering}
\begin{minipage}[t]{0.5\columnwidth}%
\begin{center}
\ 
\par\end{center}
\begin{center}
\includegraphics[width=1\textwidth]{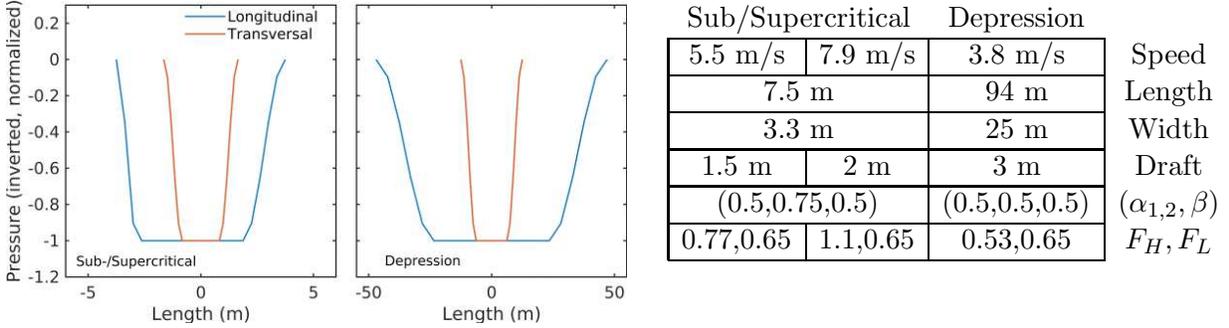}
\par\end{center}%
\end{minipage}{\small{}}%
\begin{minipage}[t]{0.5\columnwidth}%
\begin{center}
{\small{}\ }{\small\par}
\par\end{center}
\begin{center}
{\small{}}%
\begin{tabular}{c|c|c|c|c}
\multicolumn{1}{c}{} & \multicolumn{2}{c}{{\small{}Sub/Supercritical}} & \multicolumn{1}{c}{{\small{}Depression}} & \tabularnewline
\cline{2-4} \cline{3-4} \cline{4-4} 
 & {\small{}5.5 m/s} & {\small{}7.9 m/s} & {\small{}3.8 m/s} & {\small{}Speed}\tabularnewline
\cline{2-4} \cline{3-4} \cline{4-4} 
 & \multicolumn{2}{c|}{{\small{}7.5 m}} & {\small{}94 m} & {\small{}Length}\tabularnewline
\cline{2-4} \cline{3-4} \cline{4-4} 
 & \multicolumn{2}{c|}{{\small{}3.3 m}} & {\small{}25 m} & {\small{}Width}\tabularnewline
\cline{2-4} \cline{3-4} \cline{4-4} 
 & {\small{}1.5 m} & {\small{}2 m} & {\small{}3 m} & {\small{}Draft}\tabularnewline
\cline{2-4} \cline{3-4} \cline{4-4} 
 & \multicolumn{2}{c|}{{\small{}(0.5,0.75,0.5)}} & {\small{}(0.5,0.5,0.5)} & {\small{}$(\alpha_{1,2},\beta)$}\tabularnewline
\cline{2-4} \cline{3-4} \cline{4-4} 
 & {\small{}0.77,0.65} & {\small{}1.1,0.65} & {\small{}0.53,0.65} & {\small{}$F_{H},F_{L}$}\tabularnewline
\cline{2-4} \cline{3-4} \cline{4-4} 
\end{tabular}{\small\par}
\par\end{center}%
\end{minipage}{\small\par}
\par\end{centering}
\caption{Shape and navigation characteristics of boat models used in numerical simulations with realistic bathymetry. Left panels: inverted shape of the normalized pressure distribution, used as wake source for the simulations. Table: List of characteristic boat and navigation parameters.
\label{fig: boat shapes}}
\end{figure}

The distribution of mean $kh$ values over the wake population encountered in the field experiment  (figure \ref{fig: kh stats}), relevant for the shallow location of our sensors,  suggests that in the shoaling region (e.g., figure \ref{fig: sup struct}) most of the wakes are characterized by  $kh<\pi$, which is  well in the range of validity of FUNWAVE-TVD. It is not very clear, however, if this is true for their evolution across the deeper part of the channel.  To investigate this question, we conducted numerical simulations of three types of wakes (sub/super-critical and depression) using the bathymetry of the experiment site (figure \ref{fig: expmap}), and synthetic boat shapes defined by the FUNWAVE-TVD module PST-I (shape characteristics and navigation parameters given in figure \ref{fig: boat shapes}). In all simulations the ship is assumed to cruise along the center line of the channel (approximately the right-most point; figure \ref{fig: expmap}, bottom right panel). The results are illustrated using  series collected at three points in the domain, one (point A) chosen to be representative for the deeper part of the channel but distanced enough from the thalweg, and two close to the locations of the Vector ADVs (points C and D). 

\begin{figure}
\begin{centering}
\includegraphics[width=0.8\textwidth]{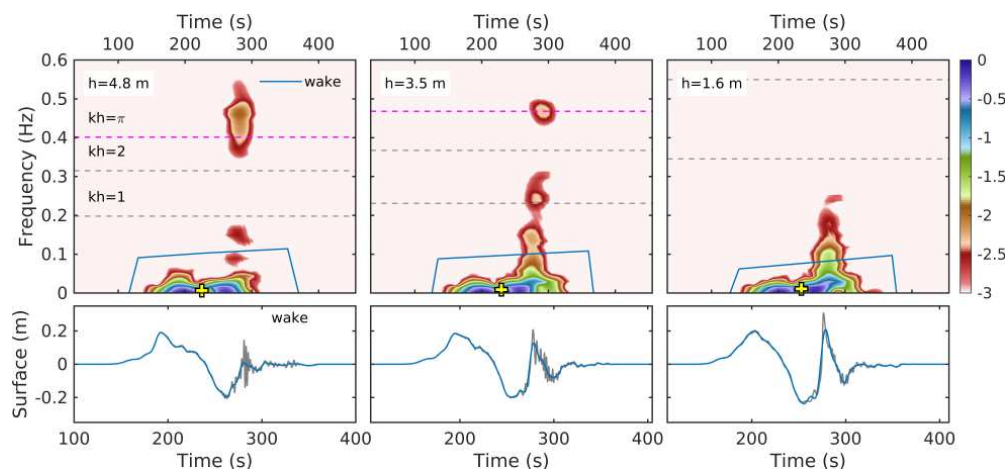}
\par\end{centering}
\caption{Example of a simulated depression (drawdown) wake created by a large ship 94-m long, 25-m wide, with a draft of 3 m, moving through the intracoastal waterway at the location of the experiment at 3.8 m/s (see figure \ref{fig: boat shapes}). Top row: normalized spectrogram ($\log_{10}$ scale, unit maximum value). The main wake is circumscribed by a polygon (blue); yellow marker indicates the center of mass of the spectrogram inside the polygon. Bottom row: free surface elevation. The filtered signal corresponding to the main wake is shown in blue, with the entire wake is given as reference in gray. The three columns show data at depths 4.8 m, 3.5 m, and 1.6 m (points A, B and C in figure \ref{fig: expmap}d). Note that the time intervals shown for the spectrogram and free surface elevation time series do not coincide -- a slight zoom in was used in the time series plotting, for clarity. 
\label{fig:  down fw}}
\end{figure}

As expected from previous discussions and examples, the  power of the depression wake (figure \ref{fig:  down fw}) is distributed time-frequency domain  well below the critical $kh=\pi$ boundary, with the exception perhaps of a high-frequency oscillation on the trailing edge of the depression, which indicates that this type of wake is likely correctly described by the model. 

\begin{figure}[h]
\begin{centering}
\includegraphics[width=0.8\textwidth]{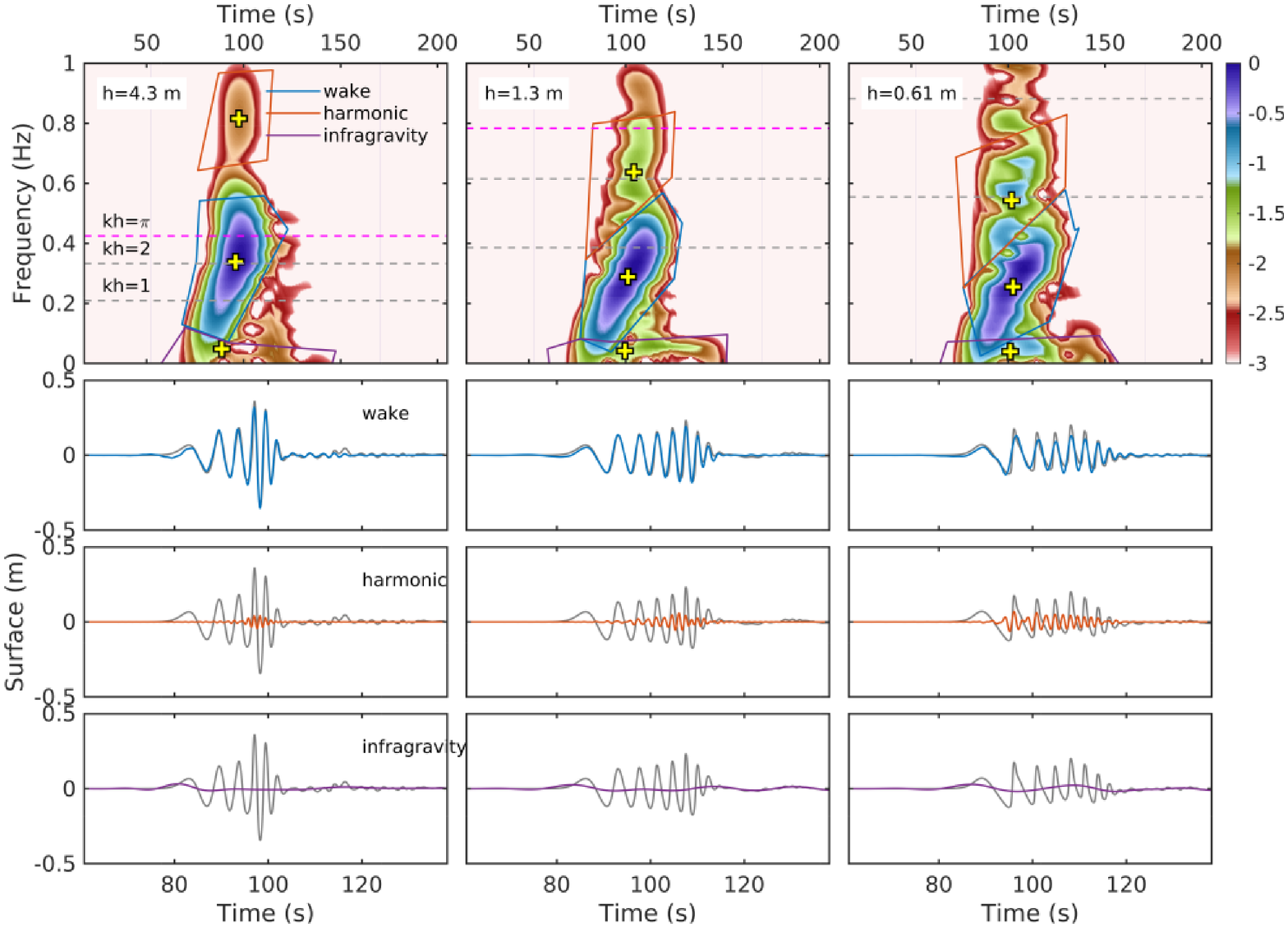}
\par\end{centering}
\caption{Example of a simulated supercritical wake created a 7.5-m long, 3.3-m wide ship with a draft of 2 m, moving through the intracoastal waterway at the location of the experiment at 7.9 m/s (see figure \ref{fig: boat shapes}). Top row: normalized spectrogram ($\log_{10}$ scale, unit maximum value). The wake components are circumscribed by polygons (blue – main wake, red – harmonic, purple – infragravity); yellow markers indicate the center of mass of the components. Bottom three rows: free surface elevation. The filtered signal corresponding to wake components is shown in the color of the polygon, with the entire wake is given as reference in gray. The three columns show data at depths 4.3 m, 1.3 m, and 0.61 m (points A, C and D in figure \ref{fig: expmap}d). Note that the time intervals shown for the spectrogram and free surface elevation time series do not coincide -- a slight zoom in was used in the time series plotting, for clarity. 
\label{fig: sup fw}}
\end{figure}

Supercritical wake simulations  (e.g., figures  \ref{fig: sup fw}) show a significant amount of energy in the problematic range of $kh\apprge\pi$ at the deepest sensor (point A, figure \ref{fig: expmap}, bottom right panel) near the 4.3-m isobath. The accumulation of energy and breaking in the $kh\ge \pi $ is easily detectable (figure \ref{fig: sup fw}, left panels),  similar to Gaussian wake tests. This suggests that frequencies above 0.45 Hz are misrepresented at this depth. However, the shallower stations shown, 1.3 and 0.6 m, that are relevant for nonlinear shoaling effects, are again within the validity range of the 2nd order (in $kh$) Boussinesq approximation, and thus are correctly described. This is also in agreement with the statistics of the mean $kh$ shown in figure \ref{fig: kh stats}. Subcritical wakes exhibit a similar behavior (not shown). 

The spectrogram filtering of the harmonic and infragravity wakes returns time series that exhibits the expected behavior for generating skewness and asymmetry in the wake, and the total wake has all the characteristics
of a shoaling wave.  The  model describes well the evolution of the nonlinear shoaling, which itself is a fundamentally more important process for sediment transport, shoreline erosion, and corresponding environmental impacts. Note that in all the wake simulations shown, the segment of the evolution that raises problems for FUNWAVE-TVD is the deeper channel, where the evolution is likely essentially linear. 

\section{Summary and discussion}\label{sec: disc}

Numerical simulations of the boat-wake climate in urbanized coastal areas play a fundamental role in understanding, forecasting, and managing the impact of the traffic on coastal wetland and reef ecosystems and sediment transport processes. Because boat wakes are highly non-stationary and inhomogeneous (i.e., temporal and spatial variations occur on scales similar to the phase scale), the class of models that can be used is restricted to phase resolving models. However, the application of phase resolving models to the wake climate problem poses a unique challenge. Because phase-resolving models require accurate, phase-scale information of forcing terms (i.e., boat and characteristics), the difficulty of accurately associating specific observed  wake field with the generating boat significantly limits the model usefulness. In specific experiments, this uncertainty may be eliminated if specific boat is used  \citep[e.g.,][]{David2017}, or if, for example, continuous video monitoring is available. However, these means are typically not available, and even if they are, they cannot help with forecasting tasks. In this study, we argue that it may be possible to circumvent this difficulty by taking into account the stochastic character of the wake climate.  

We argue that in the long time-scale relevant for understanding the environmental impacts of the wake climate, the distribution of the wake population in some relevant parameter space matters more than the details of individual wakes. Thus, an alternative approach to modeling the wake climate would be a Monte-Carlo simulation of wakes randomly chosen from the parameter space, based on the joint probability density. This hypothesis points to two directions of research: 1) the determination of the parameter space, 2) an investigation of the ability of the model of choice to represent accurately the entire wake population. 
   
The parameter space of a wake climate generated by boat traffic is likely characterized by a larger number of degrees of freedom. An exhaustive list might  include $ F_H$  and $ F_L$, but also other shape parameters, as well as navigation parameters, parameters describing wake energy, nonlinear evolution, breaking characteristics, dissipation rates, their  temporal and spatial distribution, and their relation to the local bathymetry and circulation conditions. With too many parameters, however, the problem becomes again intractable. Some research effort is necessary to identify the parameters that play an essential role. Some parameters emerge directly from the mathematical formulation oft the problem (e.g.,  the Froude numbers used here). Some other parameters might emerge from field experiments.  A given parameter would be essential if it played a dominant role in a process of interest. The list of parameters will therefore likely change depending on the purpose of the research. Detailed field  experiments will be needed to identify both the parameter space and the distribution of the wake population. It is important to note that the wake climate is fundamentally local, i.e., the results might obtained for one location might not be transferable to other locations.   

In this study, we investigated these ideas starting from a simple (crude) mathematical formulation of the wake problem, that defines two Froude numbers, $F_{H,L} $. Field observations show that these parameters provide a useful classification for wakes in the Florida Intracoastal Waterway, a boat 'highway' that supports significant commercial and recreational vessel traffic in the region. Based on these parameters, we estimated probability density functions that could be used in a future wake-climate simulation. If the environmental impact of wakes is measured by their effects on sediment transport, wake evolution in shallow water is essential. We examined nonlinear shoaling effects in wakes and showed that wake shoaling is similar to the shoaling ocean waves. This suggests that at least some of  the large volume of knowledge accumulated about sediment transport under shoaling ocean waves may apply to shoaling wakes in shallow intracoastal waterways.  

Assuming that urbanized coastal areas are essentially shallow suggests that Boussinesq models may apply to the wake climate problem. Our hypothesis that, for the wake problem,  "the distribution of the wake population matters more than the details of individual wakes" implies that a model of choice (here, FUNWAVE-TVD) should be accurate across the entire wake population. The applicability of Boussinesq models is restricted to weakly-dispersive wave fields. In general, this is not a concern, because the models are applied to systems known {\em a priori} to satisfy this condition. In the case of the wake climate, however, the dispersive character of the wakes can change across the wake population and during the wake propagation. Our numerical experiments,  covering wakes  classified using the two Froude numbers,  show that the distribution of power of a wake in time-frequency domain may exceed the limiting weak-dispersion threshold $kh\approx \pi $. In such cases, the model misrepresents the wave evolution, creating an accumulation of energy (large amplitudes of individual waves) that leads to dissipation through breaking. This problem generally arises in the deeper part of the propagation domain. It typically affects the divergent component (chirp), which naturally slides in time into high frequencies, and much less the transversal (monochromatic) component of the wake, which occupies the lowest part of the frequency band of the wake. 

Our simulations show, however, that  the problem is not as severe as it seems.  If the wave field is weakly dispersive  and the problem is completely specified  (e.g., analytical Gaussian wake cases),  the accuracy of  FUNWAVE-TVD simulations at phase resolving scale is truly excellent. Therefore, one should expect a good performance in matching realistic cases, which should cover low $F_L $ values (large container ships).  This study shows that, although the model is challenged to describe the intermediate-depth evolution of wakes produced by small slow vessels, the shallow-water nonlinear evolution, which is essential for sediment transport processes and associated erosion of coastal wetlands and reefs, is captured well {\em in all cases tested}. To correct for the intermediate-depth problem, approaches worth investigating  might be the use of higher-order Boussinesq formulations \citep[e.g.,][]{Liu2020}; or a fully-dispersive {\em linear} model and provide boundary conditions for FUNWAVE-TVD to describe the nonlinear shoaling and sediment transport processes. The former would require a higher numerical integration effort, but perhaps could be switched to a lower order dispersion approximation in shallow water. For the latter, a mathematical model more elaborate than the Gaussian model used here might work for low-energy wakes in cases where the cruise line is far enough from the shore.

\subsection*{Acknowledgments}

The field experiments discussed here were possible through the participation of University of Florida students and technicians: Deidre Herbert, Emily Astrom, Todd Van Natta, Patrick McGovern, Greg Kusel, Patrick Norby, Sofia Roman Echevarria, and Ada Bersoza as well as staff and volunteers at the GTMNERR consisting of Nicole Dix, Kenneth Rainer, Emma Hanson, Ben Mowbray, and Remo Mondazzi. We are grateful to Dr. Calantoni (Seafloor Research Branch, Naval Research Laboratory) for generously allowing us to use the q-boat for bathymetric surveys.

The work of Sheremet,  Forlini, and Qayyum was partly supported by NSF Grant No. 1737274.   

This work was sponsored in part by the National Estuarine Research Reserve System Science Collaborative, which supports collaborative research that addresses coastal management problems important to the reserves. The Science Collaborative is funded by the National Oceanic and Atmospheric Administration and managed by the University of Michigan Water Center (NAI4NOS4190145).

This research was supported in part by an appointment to the Department of Defense (DOD) Research Participation Program
administered by the Oak Ridge Institute for Science and Education (ORISE) through an interagency agreement between the U.S.
Department of Energy (DOE) and the DOD. ORISE is managed by ORAU under DOE contract number DE-SC0014664. All opinions expressed in this paper are the
author's and do not necessarily reflect the policies and views of DOD, DOE, or ORAU/ORISE.

Permission was granted by the Chief of Engineers to publish this information.

The software used in the numerical simulations is the open-source FUNWAVE-TVD model which can be found and downloaded at URL: https://fengyanshi.github.io/build/html/index.html.

Data for this research are not publicly available since they are the subject of ongoing graduate research and forms the basis of several thesis dissertations.
\bibliographystyle{plainnat}
\bibliography{wakes}

\end{document}